\documentclass{article}

\usepackage{arxiv}
\RequirePackage{amsthm,amssymb,amsmath,longtable,color, colortbl, float, nicematrix, mathtools}

\usepackage[utf8]{inputenc} 
\usepackage[T1]{fontenc}    
\usepackage{hyperref}       
\usepackage{url}            
\usepackage{booktabs}       
\usepackage{amsfonts}       
\usepackage{nicefrac}       
\usepackage{microtype}      
\usepackage{cleveref}       
\usepackage{lipsum}         
\usepackage{graphicx}
\usepackage{natbib}
\usepackage{doi}

\mathtoolsset{showonlyrefs=true}
   \usepackage{natbib}
    \bibpunct{(}{)}{;}{a}{,}{,}
    \usepackage{xpatch}
    \xpatchbibdriver{article}
    {\printfield{year}\newunit}
    {\printfield{year}}
    {}{}
\title{Non-stationarities in extreme hourly precipitation over the Piave Basin, northern Italy}

\usepackage{authblk}

\author{\large{Dáire Healy$^{1*}$, Ilaria Prosdocimi$^{1}$, and Isadora Antoniano-Villalobos$^{1}$}
}
\normalsize
\vspace{1em}
\affil{$^1$Dipartimento di Scienze Ambientali, Informatica e Statistica, Venezia, Italia}
\affil{$^*$\texttt{\url{daire.healy@ucd.ie}}}

\date{}
\begin{document}
\maketitle
\begin{abstract}
We study the spatio-temporal features of extremal sub-daily precipitation data over the Piave river basin in northeast Italy using a rich database of observed hourly rainfall. Empirical evidence suggests that both the marginal and dependence structures for extreme precipitation in the area exhibit seasonal patterns, and spatial dependence appears to weaken as events become more extreme. We investigate factors affecting the marginal distributions, the spatial dependence and the interplay between them. Capturing these features is essential to provide a realistic description of extreme precipitation processes in order to better estimate their associated risks. With this aim, we identify various climatic covariates at different spatio-temporal scales and explore their usefulness. We go beyond existing literature by investigating and comparing the performance of recently proposed covariate-dependent models for both the marginal and dependence structures of extremes. 
Furthermore, a flexible max-id model, which encompasses both asymptotic dependence and independence, is used to learn about the spatio-temporal variability of rainfall processes at extreme levels. We find that modelling non-stationarity only at the marginal level does not fully capture the variability of precipitation extremes, and that it is important to also capture the seasonal variation of extremal dependence.
\end{abstract}


\section{Introduction}

Extreme precipitation events in northern Italy have become an increasing concern due to their devastating impacts, including significant flooding and landslides, which have major environmental and socioeconomic consequences \citep{Spano2020}. Recent instances of such events underscore the urgency of this issue \citep{Cassola2023, Donnini2023, Dorrington2024}. Detecting trends in extreme precipitation at a national scale remains challenging in Italy, likely due to the country's diverse climate, complex terrain, and spatio-temporal dynamics of precipitation processes \citep{Libertino2019}. Inconsistent patterns across Italy \citep{Brugnara2019}, suggest that national-scale analyses, based on coarse resolution datasets, may obscure local dynamics caused by the country's high climatic and geographic heterogeneity. Extreme precipitation in northern Italy is particularly challenging to model due to its unique geographical and climatological setting, influenced by the Alps.  Heavy precipitation trends in northern Italy show heterogeneous patterns, highlighting the necessity for localised studies within the region \citep{Isotta2014, Crespi2018,Libertino2019}. The literature calls for more nuanced local studies of extreme precipitation trends to effectively inform local authorities and address regional vulnerabilities. A localised understanding of precipitation extremes is critical for water management, infrastructure development, and over all risk mitigation \citep{Rahat2024}.

To estimate the magnitude of rare precipitation events, extreme value theory (EVT) is routinely applied \citep{Katz2002, Coles2003, Papalexiou2013}. EVT provides a theoretically justified approach to identify the limiting distribution for extremes defined as either block maxima, i.e., the largest value within blocks of a pre-specified size, or peaks over threshold, i.e., all exceedances larger than a certain high threshold. Using probability theory, it is possible to derive the limiting distributions of these extremes: the Generalised Extreme Value (GEV) distribution for block maxima, and the Generalised Pareto (GP) distribution for threshold exceedances \citep[see][]{Coles2001}. EVT has furthermore been adapted to investigate changes in the distribution of extreme rainfall linked to climate change and other climatic drivers \citep{Jayaweera2024}. Several studies propose methods to reparametrise the GEV and the GP distributions to allow non-stationarity in the extremal models \citep{Eastoe2009b, prosdocimi2021parametrisation}, enabling the practitioner to capture trends in the intensity and frequency of extreme events. In northern Italy, the use of EVT has significantly enriched our knowledge of the behaviour and the associated risks of extreme precipitation. For example, \cite{Crespi2018} identified long-term trends in the region, where some areas have seen increased intensities and frequencies of heavy rainfall, while others have remained relatively stable. \cite{Gentilucci2023} emphasises the inadequacy of stationary models of extreme precipitation given the increased variability of extreme rainfall over Italy in recent years. These findings highlight the utility of EVT models as a crucial tool for characterising changes in extreme precipitation patterns.

Univariate models fail to capture the inherent spatial dependencies of extremal processes. A natural multivariate extension of the univariate block maxima approach is the max-stable process, which can be seen as an infinite-dimensional analogue of the GEV distribution, developed to describe the limiting behaviour of component-wise maxima in a spatial setting \citep{deHaan1984, Davison2012}. Max-stable processes provide a popular and widely used framework for modelling the joint extremes of precipitation over space. A major concern with the use of max-stable processes is that they assume a very strong form of extremal dependence, known as asymptotic dependence. This assumption should not be made a priori, since it oversimplifies the dependence structure and may lead to the incorrect estimation of extreme events. It is often observed that environmental processes such as precipitation are not asymptotically dependent and become site-wise independent in the uppermost tail. To account for this, max-id models have been proposed as a flexible alternative to max-stable models. They are able to capture asymptotic independence, i.e., a weakening, eventually vanishing, extremal dependence at increasing quantiles \citep{Padoan2013, Bopp2021, Huser2021, Zhong2024}. 

Our analysis entails a two-level modelling procedure. Firstly, we model the marginal distributions of the process using a covariate-dependent GEV distribution, accounting for seasonality, temporal non-stationarity and other geophysical influences acting on the extremal process. Secondly, we account for the extremal dependence of the process through a spatial max-id extreme value model. We enrich the original max-id process model by introducing covariate-dependent parameters to deal with the seasonal non-stationary behaviour of the extremal dependence. We find that this more complex model more accurately captures the spatial and temporal structure of the extreme precipitation data.

The paper is organised as follows. Section~\ref{sec:data} describes the observational data and study region. Section~\ref{sec:explor} provides an exploratory analysis to identify possible non-stationarities in the data, both marginally and spatially. Section~\ref{sec:covars} introduces the covariates used to account for the identified non-stationarities. Section~\ref{sec:methodology} outlines the marginal and dependence models. In Section~\ref{sec:results}, we present and discuss the results of our analysis. Conclusions and broader discussion are provided in Section~\ref{sec:discussion}. 

\section{Data}\label{sec:data}
\begin{figure}[ht]
    \centering
    \includegraphics[width=0.9\linewidth]{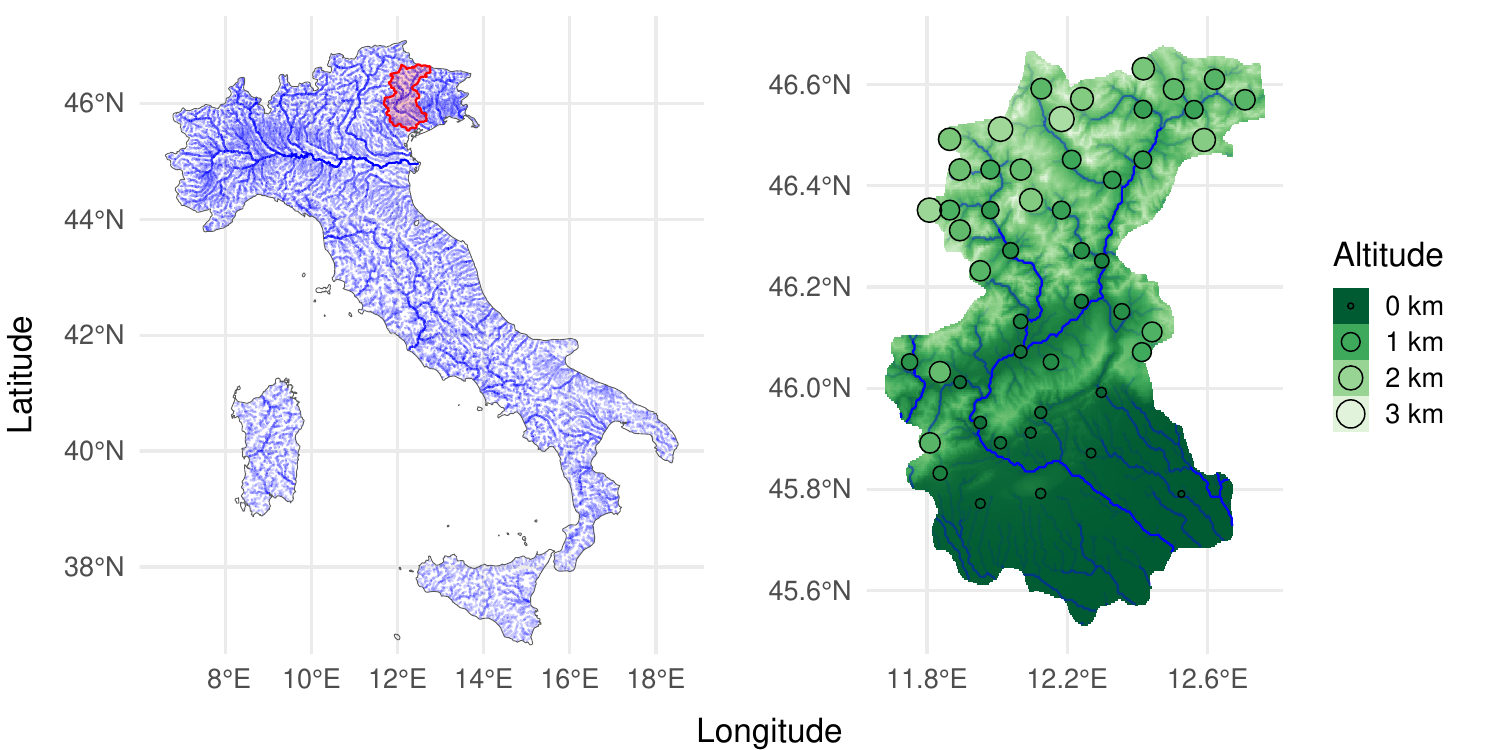}
    \caption{(Left) Plot of river basins in Italy, with the study region of Treviso and Belluno provinces highlighted in red. (Right) A focus on the Piave river basin (region highlighted in red on the left-hand plot) with rivers and tributaries shown. Observation sites are plotted with size and colour corresponding to the altitude of the site.}
    \label{fig:river_basin_alt_map}
\end{figure}

We model the spatio-temporal behaviour of monthly maxima of hourly precipitation over the Piave basin, in north-east Italy. The study domain is highlighted in Figure~\ref{fig:river_basin_alt_map}. Monthly maxima are derived from a larger dataset of hourly precipitation recorded by the local environment monitoring agencies (Agenzia Regionale per la Prevenzione e Protezione Ambientale del Veneto, ARPAV\footnote{\url{https://www.arpa.veneto.it/dati-ambientali/dati-storici}}). The initial dataset consists of records from 65 observational sites between 1990 and 2023, with an average span of 27.8 years. In keeping with standard statistical modelling of precipitation processes, we define the day as starting from 09:00 AM. We remove stations with a substantial number of missing values; specifically, 4 sites with more than 20\% missing data are excluded. To avoid deriving inaccurate or biased maxima due to gaps in the data, we omit any month which does not have near-complete observational coverage. Thus, we omit months containing fewer than 80\% of days observed for each site separately. In this way, we remove approximately 1.23\% of months. In order to mitigate numerical issues and further lessen the computational cost and complexity of our spatial model (see Section~\ref{sec:spatial_models}), we exclude sites that are located within 5 km of each other. These sites are kept as a test-set for out-of-sample verification. In each case, we keep the site with the longer observation record, resulting in a total of 49 sites to be considered in the analysis. The altitude and spatial distribution of the modelled sites can be seen on the right-hand panel of Figure~\ref{fig:river_basin_alt_map}.

Let $X(t, \boldsymbol{s})$ be the observed precipitation at time, $t \in \mathcal{T}$, and site, $\boldsymbol{s}\in \mathcal{S} \subseteq \mathbb{R}^2$. The set $\mathcal{T}$ consist of hourly timestamps $t_1, t_2, \dots \in \mathcal{T}$, where $t_i$ is the timestamp to hourly resolution of the $i$-th observation. The set $\mathcal{S}$ comprises the locations $\boldsymbol{s}_1, \boldsymbol{s}_2, \dots \in \mathcal{S}$, where $\boldsymbol{s}_i$ is a 2D spatial coordinate vector, i.e., $\boldsymbol{s}_i = (\text{long}_i, \text{lat}_i)$. Monthly maxima of hourly precipitation data are taken for each site separately. We define $M(m_t, y_t, \boldsymbol{s})$ as the maximum precipitation at location $\boldsymbol{s}$, in month $m_t$, year $y_t$, where $m_t \in \{1,2,\dots, 12\}$ is the month in which the observation at time $t$ occurred, and $y_t \in \{1990, 1991, \dots, 2023\}$ the corresponding year.

\section{Exploratory analysis}\label{sec:explor}
In this section, we conduct an exploratory analysis to examine spatial and temporal features of monthly precipitation maxima over the Piave basin. We explore non-stationarities in both the marginal and spatial dependence structure of the process. This analysis aims to inform our set of potential physical covariates and model parameterisation to better capture the marginal and extremal dependence structures of the data. 

\subsection{Long-term temporal non-stationarity} \label{sec:expl_long_term_variations}
\begin{figure}[ht]
    \centering
    \includegraphics[width=0.8\linewidth]{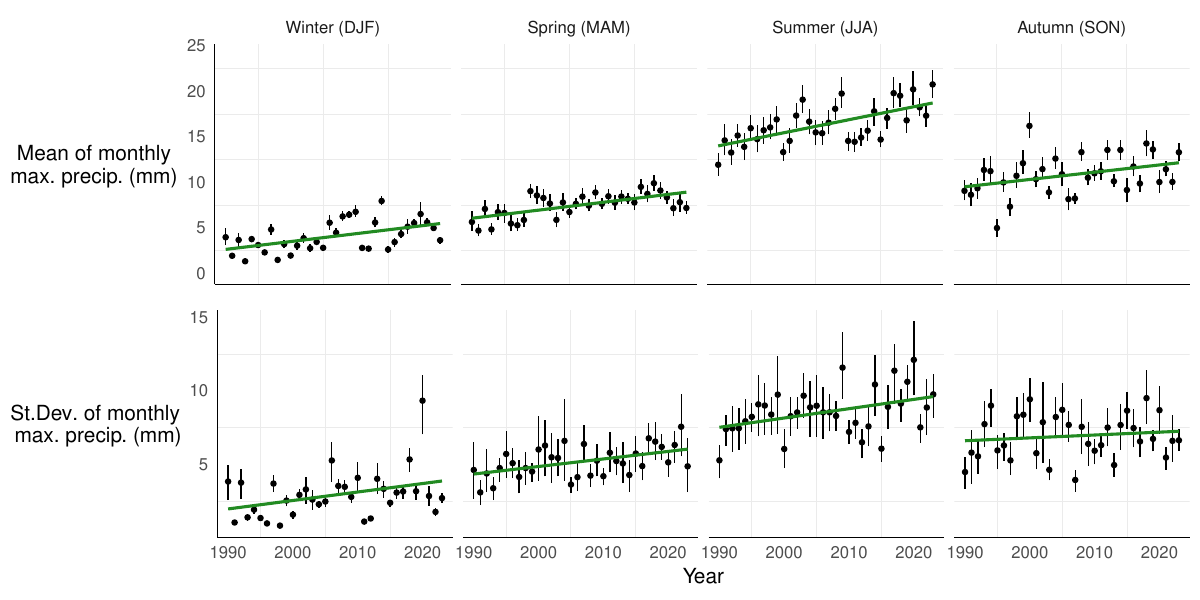}
    \caption{Empirical long-term trends in summary statistics of monthly block maxima of hourly precipitation for each season. Simple regression lines are shown in green as a visual aid.  
    (Top row) Average monthly maxima over the study domain for each year, for each season. (Bottom row) Standard deviation of monthly maxima over the study domain was calculated for each season and each year separately. Displayed uncertainties are based on 500 bootstrap samples within each season, within each year.}
    \label{fig:exploritory_mean_sd}
\end{figure}

Understanding long-term trends in extreme precipitation is crucial for capturing how this process is changing in response to a warming climate. Rising global temperatures are expected to alter atmospheric moisture content and weather patterns, and there is increasing evidence that a warmer climate is positively correlated with rainfall frequency and intensity \citep{Barbero2017, Ali2022}. However, relatively little is known about how sub-daily \textit{extreme} precipitation is changing in response to climate change, especially at local scales. In Figure~\ref{fig:exploritory_mean_sd} we plot the average and standard deviation (calculated over space) of monthly maxima of hourly precipitation for each year and each season separately. We also plot the 95\% uncertainty associated with each estimate using a simple bootstrap procedure within each season and year. In each season, for both mean and standard deviation, we see an increase over time, suggesting that extreme precipitation has become more intense and variable over the study period. To investigate this further, we incorporate long-term temporal non-stationarity in our marginal model of the process, after accounting for other, perhaps obfuscating or enhancing signals (such as seasonality). We use temperature anomalies as a predictor (see Section \ref{sec:cov_marg}). Temperature anomalies have an increasing long-term trend and have an interpretation in physical terms, unlike time. 

\subsection{Seasonality}\label{sec:expl_seasonality}

We anticipate that seasonality plays a critical role in shaping the marginal distribution and potentially the extremal dependence structure of extreme precipitation events over the study region \citep{Cabral2020, Jurado2023}. Different weather systems dominate throughout the year which influence the frequency, intensity, and spatial extent of extreme precipitation. Ignoring these variations can mask other trends in the underlying processes. 

Several studies have investigated seasonality of daily precipitation extremes at a global scale. Climate model projections suggest that both the magnitude and timing of daily extreme precipitation are expected to change, with some regions experiencing earlier extremes and others seeing delays \citep{Treppiedi2025}. These seasonal shifts in daily precipitation are not uniform, with some regions showing little or inconsistent change \citep{Marelle2018, Zhu2025}. Previous work identifies strong regional heterogeneity in seasonal dynamics of precipitation extremes, highlighting the importance of regional analyses. Relatively little work has been done on seasonal analysis of sub-daily precipitation extremes.

\begin{figure}[ht]
    \centering    \includegraphics[width=0.8\linewidth]{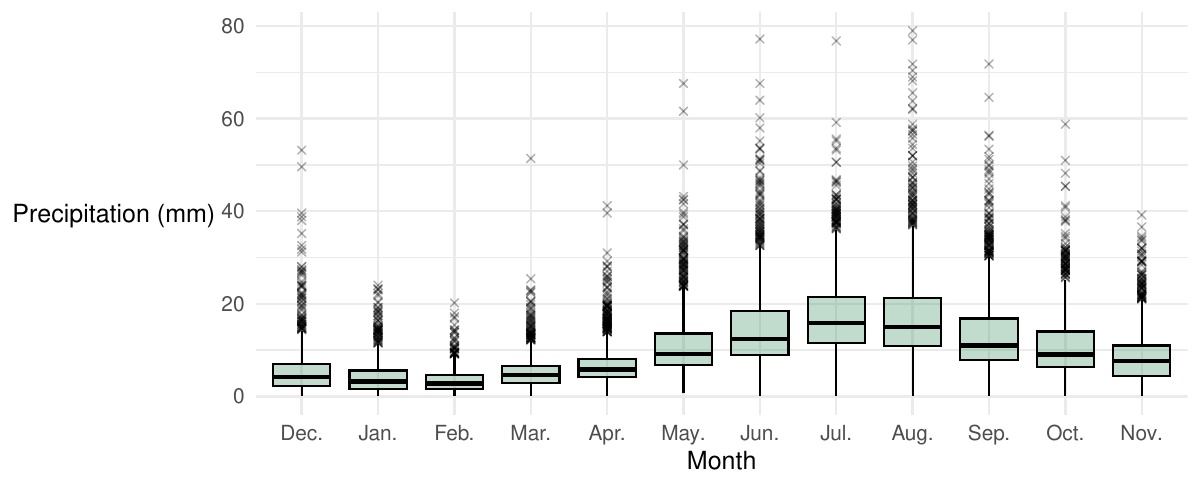}
    \caption{Distribution of monthly maxima of hourly precipitation data for each month.}
    \label{fig:mean_monthly_max_obs}
\end{figure}

Figure~\ref{fig:mean_monthly_max_obs} illustrates a seasonal trend, showing that the most extreme observations occur in the summer months (June, July and August) and the least extreme hourly precipitation events occur in late winter or early spring (January, February and March). To account for the observed seasonality, we include month as a predictor in the model. 

\subsection{Spatial variability}\label{sec:exp_spat_var}
\begin{figure}[ht]
    \centering
    \includegraphics[width=0.9\linewidth]{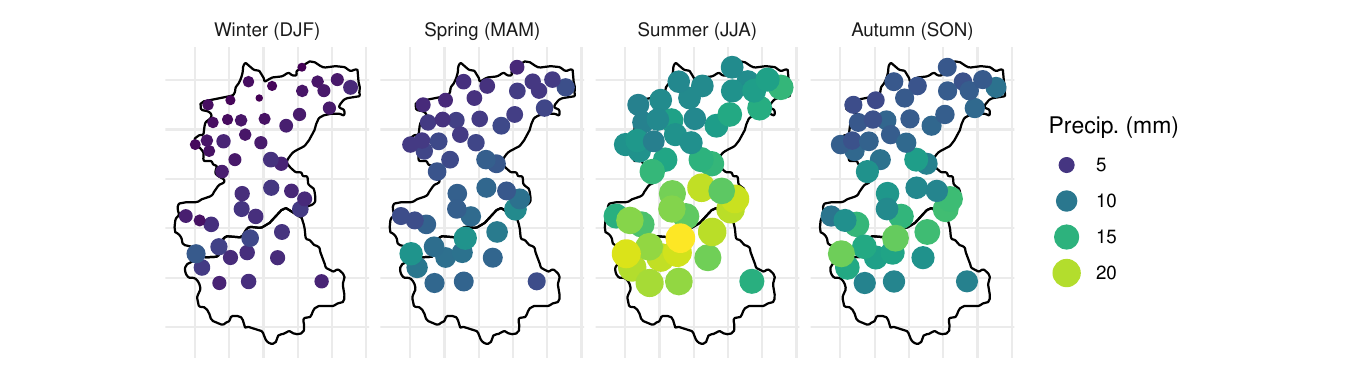}
    \caption{Median monthly maxima at each site for each season winter-autumn (left to right).}
    \label{fig:seas_spatial_var}
\end{figure}

Northeastern Italy is a region marked by complex climatic and topographic variability. The terrain's heterogeneity, including substantial variations in elevation, slope, and aspect, results in complex orographic lifting mechanisms producing highly variable precipitation processes. Further complications are introduced through interactions between orographic lifting, local convective storms, and large-scale precipitation systems \citep{Formetta2022}. These effects often make extreme precipitation highly localised, sometimes with marked differences over short distances \citep{Mazzoglio2025}. For example, \cite{Isotta2014} identified regions south of the Alps that receive considerably higher rainfall due to orographic influences. Figure~\ref{fig:seas_spatial_var} illustrates the spatial variability in our data, and how spatial variability changes with season. The figure highlights the dramatic increase in precipitation magnitude in the pre-Alpine region. The complex geophysical features make it a particularly challenging region to perform statistical modelling of precipitation, even in the presence of a large set of predictors. To achieve parsimony, we exploit a high-resolution hourly climate model, which mostly captures the fine-scale spatial variability of precipitation (see \ref{sec:cov_marg}) to construct a low-dimensional set of predictors. Furthermore, we also include coastal distance as a predictor to capture any residual effect linked to different processes influenced by the sea. 

\subsection{Extremal dependence and non-stationarity therein}\label{sec:asymp_dep}
We conclude the exploratory analysis by examining the extremal dependence of hourly precipitation between observations from stations at various distances from each other. Given two random variables $X_1$ and $X_2$, their tail coefficient $\chi$ is an extremal dependence index defined as
\begin{align}
    \chi(X_1,X_2) = \lim_{z\to1^-} {Pr} \left\{ F_1(X_1) > z |  F_2(X_2) > z\right\},
\end{align}
where $F_i$ denotes the distribution function of $X_i$ \citep{Coles1999}. 
An estimator of the tail coefficient of $\chi$ at quantile $q$ is given by
\begin{equation}  
\label{eq:chi_lim2}
    \chi_q(X_1,X_2) =  {Pr} \left\{ {F}_1(X_1) > q  | {F}_2(X_2) > q\right\},
\end{equation}
which can be estimated empirically for $0<q<1$.

Calculating $\chi_q(X_1, X_2)$ over a sequence of high quantiles provides insight into the extremal dependence regime of the process. If $\chi(X_1, X_2) > 0$, or in practice, if $\chi_q(X_1,X_2) \to k >0$ as $q \to 1$, the variables are judged to be asymptotically dependent. If $\chi(X_1, X_2) = 0$, or in practice, if $ \chi_q(X_1, X_2) \to 0$ as $q \to 1$, the variables are interpreted as asymptotically independent. 
Identifying whether the data exhibits asymptotic dependence or asymptotic independence is crucial as it directly informs the appropriate modelling approaches. 

\begin{figure}[ht]
    \centering
    \includegraphics[width=\linewidth]{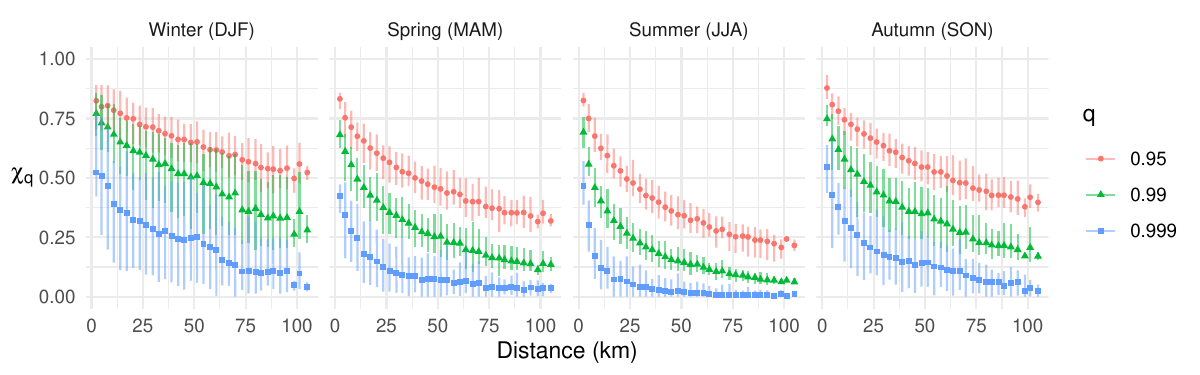}
    \caption{Estimates of extremal dependence coefficient, $\chi_q$, for each season at a range of quantiles $q = [0.95, 0.99, 0.999]$.}
    \label{fig:chi_estimates}
\end{figure}

For each pair of synoptic sites in our data, we estimate the tail coefficient, taking empirical distribution functions for each site and using \eqref{eq:chi_lim2} for $q \in \{ 0.95, 0.99, 
0.999 \}$. We do this for each season separately, winter (December, January, February), spring (March, April, May), summer (June, July, August) and autumn (September, October, November). We group the pairwise estimates into 35 bins based on the distance between the pairs and plot the point estimate of $\chi_q$, corresponding to the mean value of all $\chi_q$ estimates in that bin, along with the 95\% uncertainty, corresponding to the 95\% quantiles of all $\chi_q$ estimates in that bin. These empirical estimates are plotted against the distance between sites in Figure~\ref{fig:chi_estimates}. For all seasons, the strength of the extremal dependence diminishes with distance. However, we can see a clear seasonal pattern in the extremal dependence structure of the process. This exploratory analysis suggests that extreme dependence is strongest during the winter months and persists over long distances (indicating more widespread extreme events). In contrast, during the summer months, extreme dependence tapers off at shorter distances (indicating more localised extreme events). Critically, in each season, we see extreme dependence weakening with quantile level $q$, suggesting the data exhibits asymptotic independence, and so a model which captures this is necessary.  

A major difficulty in modelling extremal dependence is the problem of distinguishing whether non-stationarity in the dependence structure is an intrinsic property of the process or, rather, simply an artefact of non-stationarity in the marginals \citep{Kakampakou2024}. If seasonality in the margins explains away the seasonal pattern in extremal dependence, we can conclude that dependence remains stationary after accounting for seasonality in the margins. Although there are numerous case studies regarding the treatment of non-stationarity in extreme environmental processes, it is usually treated at the marginal level. Non-stationarity in the marginal distributions can be addressed by incorporating temporal or climatic covariates \citep{Coles2001, Eastoe2009}. In contrast, modelling non-stationarity in the dependence structure itself requires more sophisticated frameworks, which allow for temporal or spatial variations in the dependence parameters \cite{Zhong2022}. Our proposed model incorporates seasonality in the marginal distributions, allowing us to assess whether this adjustment sufficiently accounts for all observed seasonal variations. Therefore, we fit both stationary and seasonally varying dependence models after removing marginal non-stationarity and compare the results to determine whether a flexible marginal model suffices to describe the seasonal variation in extremal dependence. This approach avoids the assumption commonly made in extreme value analysis, that even though marginals may be non-stationary, dependence structures remain invariant over time.

\section{Covariates}\label{sec:covars}
We aim to account for several sources of non-stationarity; namely, long-term temporal non-stationarity, seasonality and spatial non-stationarity as detailed in Section~\ref{sec:explor}. In the following subsections, we describe the covariates explored to account for each of these features in the data. In Section~\ref{sec:cov_marg}, we detail the covariates explored to account for the spatio-temporal non-stationarity of monthly maxima of precipitation extremes. In Section~\ref{sec:cov_dep}, we present a covariate to account for seasonality in the extremal dependence of component-wise monthly maxima. Additionally, we discuss an adjusted distance metric to better account for topographical influences on spatial extremal dependence. 

\subsection{Physical covariates for the marginal model}\label{sec:cov_marg}
\begin{figure}[ht]
    \centering
    \includegraphics[width=0.8\linewidth]{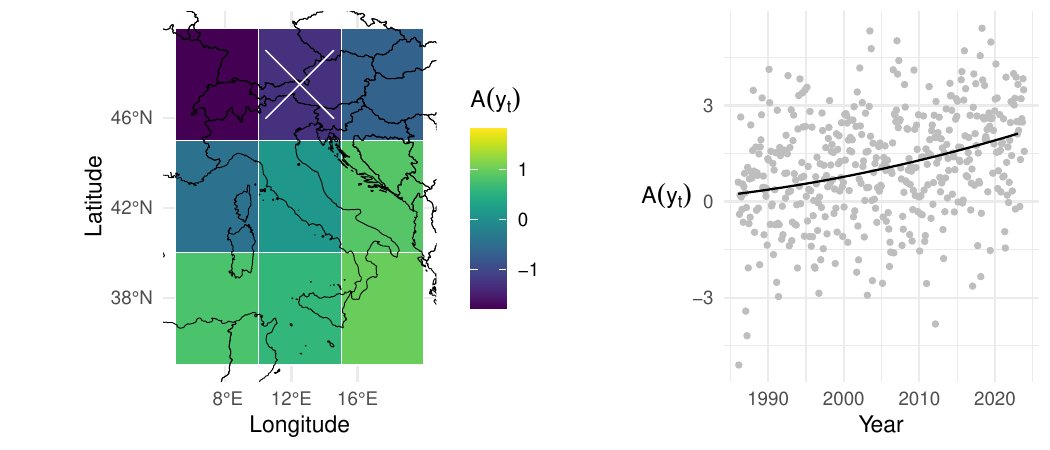}
    \caption{An event from the HadCRUT5 grid over Italy in July 2000. The HADCRUT5 grid box for northeastern Italy is marked by a $\times$ (left); LOESS smoothed north-east Italian temperature anomalies, with raw HadCRUT5 monthly temperature anomaly data shown in lighter points (right).}
    \label{fig:hadcrut_data}
\end{figure}

We choose temperature anomalies as a covariate to capture the long-term temporal non-stationarity in precipitation extremes identified in Section~\ref{sec:expl_long_term_variations}. Changes in temperature are a physically meaningful and accurate indicator of climate change. We obtain temperature anomalies from the HadCRUT5 dataset \citep{Morice2021}. The HadCRUT5 dataset provides monthly temperature anomalies on a global grid with a spatial resolution of 5×5 degrees, a section of which can be seen in the left-hand panel of Figure~\ref{fig:hadcrut_data}. The data are produced by merging a land temperature anomaly dataset, CRUTEM5, with a Sea Surface Temperature (SST) anomaly dataset, HadSST4. SSTs greatly affect atmospheric patterns and atmospheric moisture content, an informative aspect in our analysis of extremal precipitation. The HadCRUT5 dataset also captures large-scale climatic oscillations, such as the North Atlantic Oscillation (NAO) and the Mediterranean Oscillation (MO), which are known to affect weather patterns over northern Italy \citep{Baronetti2022}. This dataset parsimoniously captures complex interactions between the ocean, land and atmosphere, whose interconnectivity is crucial for understanding precipitation dynamics across time. To derive a temporal covariate from the HadCRUT5 dataset, we take the time series of monthly temperature anomalies from the closest grid point to our study domain (marked with a `$\times$' in the left-hand plot of Figure~\ref{fig:hadcrut_data}) and perform a LOESS smoothing over years, denoting smoothed temperature anomalies during year $y_t$ as $A(y_t)$. For the LOESS smoothing, we use the base \texttt{R} implementation, with the default settings of a span of $0.75$ and a degree of $2$. The smoothed temperature anomalies, $A(y_t)$, can be seen in the right-hand panel of Figure~\ref{fig:hadcrut_data} along with unsmoothed monthly anomalies. 

To account for spatial non-stationarity (Section~\ref{sec:exp_spat_var}), we derive covariates from high-resolution hourly climate model outputs. To this end, we use the Very High-Resolution PROjections for ITaly (\texttt{VHR-PRO\_IT}) dataset \citep{raffa2023}. This dataset is derived using dynamical downscaling of Global Climate Models configured with \textit{historical experiment}, reflecting observed natural and anthropogenic atmospheric composition for the period 1981–2005. The downscaled projection provides simulated data of hourly weather conditions over Italy from 1981 to 2020 on a 2.2$^\circ$  grid, which is on the Convection Permitting Scale. Dynamical downscaling involves using an ensemble of a Regional Climate Model (RCM) and Global Climate Model (GCM) pairs to simulate climatic variables at a higher resolution. The RCM helps encode very detailed information about local topography, land use, and complex weather patterns. Critically, the RCM provides information that is not observable in the weather station grid resolution. Ultimately, the inclusion of covariates derived from this dataset enriches the spatial information of the model, capturing fine-scale topographic features as well as the influence of large-scale climatic factors on local precipitation extremes. 

\begin{figure}[!h]
    \centering
    \includegraphics[width = 0.8\linewidth]{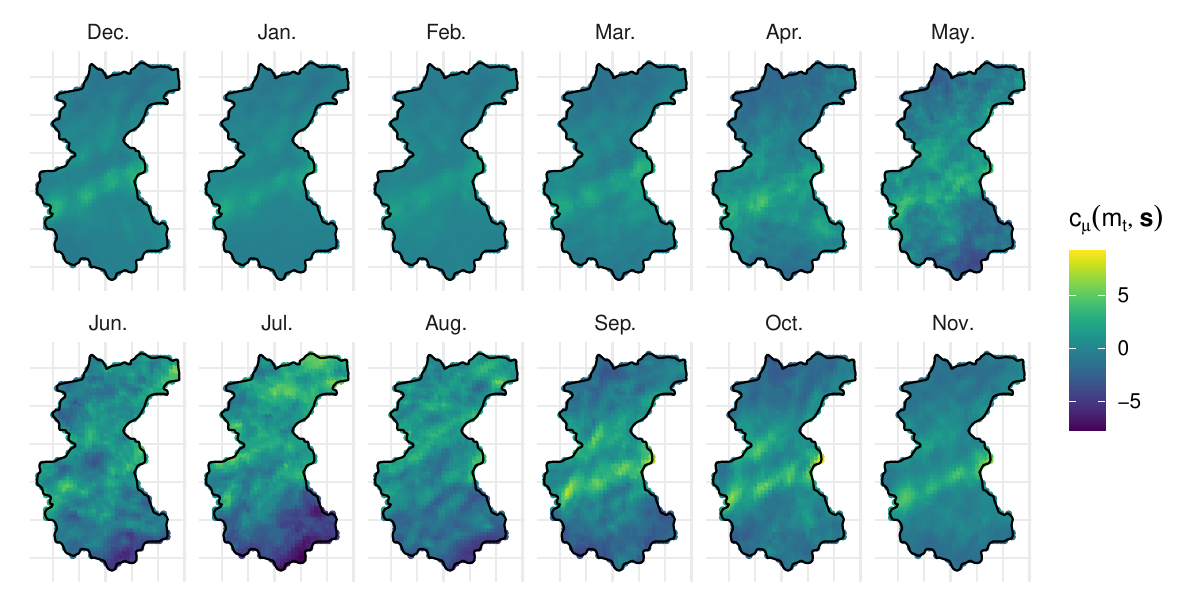}
    \caption{Centered mean monthly maxima $c_\mu(m_t, \boldsymbol{s})$, of hourly precipitation from \texttt{VHR-PRO\_IT} dataset.}
    \label{fig:clim_mean_max}
\end{figure}

For the grid points of the \texttt{VHR-PRO\_IT} dataset which overlap our study domain, we take monthly maxima for each grid point separately. We derive a spatio-seasonal covariate, capturing the unique spatial characteristics of extreme precipitation in each month. This approach reflects the different patterns of extreme precipitation that can occur throughout the year, recognising that the weather systems which drive extreme precipitation are not uniform across seasons, as observed in Section~\ref{sec:exp_spat_var}. Let $M_{\text{c}}(m_t, y_t, \boldsymbol{s})$ denote the monthly maxima of climate model data for month $m_t$, year $y_t$ and location $\boldsymbol{s}$, then 
\begin{equation}
    \tilde c_\mu(m_t, \boldsymbol{s}) = \frac{1}{n}\sum_{y_t = 1981 }^{2005} M_{\text{c}}(m_t, y_t, \boldsymbol{s}),
\end{equation}
denotes the mean monthly maxima across years. The covariate, $\tilde c_\mu(m_t, \boldsymbol{s}) $, describes typical extremal behaviour of each location and captures information related to orography and the local geography. To avoid inheriting information about the magnitudes of precipitation maxima from the climate model (which we rely entirely on observed records to inform), we centre the covariate $\tilde c_\mu(m_t, \boldsymbol{s})$. For each location, $\boldsymbol{s}$, and month, $m_t$, we subtract the spatial average from $\tilde c_\mu(m_t, \boldsymbol{s})$. Our proposed covariate, i.e., the centred version of $\tilde c_\mu(m_t, \boldsymbol{s})$, is defined as
\begin{equation}
 c_\mu(m_t, \boldsymbol{s}) = \tilde c_\mu(m_t, \boldsymbol{s}) - \frac{1}{|\mathcal{S}_c|}\sum_{\boldsymbol{s}\in\mathcal{S}_c}\tilde c_\mu(m_t, \boldsymbol{s}),
\end{equation}
where $\mathcal{S}_c$ is the set of spatial locations of the climate model output. The centred mean monthly maxima, $c_\mu(m_t,\boldsymbol{s})$ can be seen in Figure~\ref{fig:clim_mean_max}. This covariate parsimoniously describes the relative behaviour of monthly maxima at each location with respect to the others, for each month separately. 

\begin{figure}[h]
    \centering
    \includegraphics[width=0.25\linewidth]{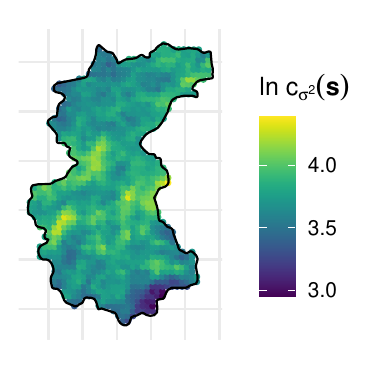}
    \caption{Log variance of mean monthly maxima $\ln c_{\sigma^2}(\boldsymbol{s})$, of hourly precipitation from \texttt{VHR-PRO\_IT}
dataset}
    \label{fig:clim_var_max}
\end{figure}

Additionally, to leverage information about the spatial variability of extreme precipitation over the region from the climate model output, we calculate the log variance for each site of monthly maxima, $\ln c_{\sigma^2}(\boldsymbol{s})$, over all months. This covariate can be seen in Figure~\ref{fig:clim_var_max}. The spatial covariates capture the dominating influence of the Prealps, the first orographic obstacle to the atmospheric systems moving northward in the region. This region typically receives markedly more precipitation \citep{Isotta2014, Formetta2024}, which is reflected by the band of higher mean monthly maxima, $c_\mu(m_t,\boldsymbol{s})$, and variance, $\ln c_{\sigma^2}(\boldsymbol{s})$, values visible across the centre of the maps. 

\subsection{Physical covariates for extremal dependence structure}\label{sec:cov_dep}
\begin{figure}[h]
    \centering
    \includegraphics[width=0.5\linewidth]{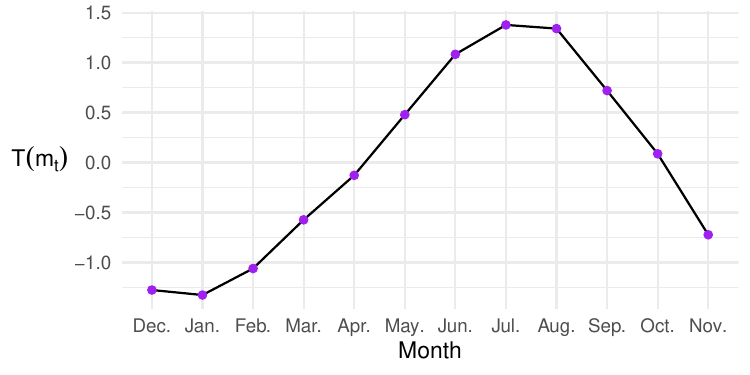}
    \caption{Standardised mean of max-daily temperature $T(m_t)$, taken as an average over the study domain for each month, $m_t$. }
    \label{fig:mean_max_daily_temp}
\end{figure}

We develop a covariate-dependent extremal dependence model which can capture observed seasonal trends identified through exploratory analysis in Section~\ref{sec:asymp_dep}. We take the standardised mean of observed maximum daily temperature over observed stations, denoted for month $m_t$ as $T(m_t)$, as a covariate. This covariate, as shown in Figure~\ref{fig:mean_max_daily_temp}, is appealing due to its smooth and sinusoidal behaviour, reflecting the anticipated seasonality of extremal dependence. The choice of mean monthly temperature reflects environmental conditions that drive different precipitation processes, which are dominant at different periods of the year, where warmer temperatures during summer months drive localised convective storms, and cooler temperatures in winter months result in stratiform precipitation \citep{Miglietta2022}.

Topographic features strongly influence precipitation distribution due to orographic lifting and shadowing effects. Euclidean distance does not account for physical obstacles over which atmospheric processes operate. A topographically adjusted distance provides a more realistic spatial separation for the observation sites. Incorporating topographic distance into the spatial dependence structure improves the ability of the model to capture localised precipitation processes. Regions separated by a mountain ridge may have a weaker extremal dependence than suggested by Euclidean measures due to the physical barrier, while valleys may be more strongly connected, which is more accurately reflected by topographic distance. Rather than taking the traditional Euclidean (or Haversine) distance between two sites, we use a topographically adjusted distance, correcting straight-line distance for topographic slope effects, to help account for the large spatial variability in the region. We calculate the topographically corrected distance, $d \in \mathbb{R}_+$ (in km), between two sites using the \texttt{R} package \texttt{spatialEco} \citep{Evans2023}. 

\section{Methodology}\label{sec:methodology}
In this section, a detailed description of the methodology used to model extreme precipitation is set out. Our approach comprises a two-stage modelling procedure. Firstly, in Section~\ref{sec:marg_mod}, we provide a detailed characterisation of the marginal distribution of the process, accounting for all non-stationarities identified in Section~\ref{sec:explor}. Secondly, in Section~\ref{sec:spatial_models} we account for the spatial dependence structure of the process, identifying seasonality in the extremal dependence structure not captured by the marginal model. Together, these components provide a complete stochastic characterisation of the extremal process.

\subsection{Marginal model}\label{sec:marg_mod}
There are two predominant methods for modelling univariate extreme value processes in the literature, namely the block maxima approach (BM) \citep{Gumbel1958} and the threshold exceedance approach, also referred to as the peak over threshold approach (\citealp{Todorovic1970}, \citealp{Pickands1975}). In this study, we use the block maxima approach to model extreme hourly precipitation data. For block maxima of a sequence of an \textit{iid} random variable, the limiting distribution (where the limit is taken for the block size going to infinity) can be shown to be the Generalised Extreme Value (GEV) distribution with CDF 
\begin{equation}\label{eq:gev}
    G(x)=\exp \left\{-\left[1+\xi\left(\frac{x-\mu}{\sigma}\right)\right]_+^{-1 / \xi}\right\}.
\end{equation}
The GEV distribution is parameterised by the location parameter, $\mu\in \mathbb{R}$, the scale parameter, $\sigma>0$, and the shape parameter, $\xi\in \mathbb{R}$. The three possible upper-tail characteristics of the GEV are recovered by specifying $\xi$, the shape parameter, i.e., $\xi = 0$ gives the Gumbel, $\xi > 0$ gives the Fréchet, and $\xi < 0$ gives the reversed Weibull distribution \citep{Coles2001}.

To account for spatial and temporal non-stationarity in the marginal distribution of the process $M(m_t, y_t, \boldsymbol{s})$, we explore a re-parameterisation of the GEV distribution in Equation~\eqref{eq:gev} that incorporates the covariates detailed in Section~\ref{sec:covars}. Specifically, we use a Generalised Additive Model (GAM) framework for the GEV parameters, implemented via the \texttt{R} package \texttt{evgam} \citep{Youngman2020}, which allows for both linear and smooth additive relationships between covariates and GEV parameters. The use of GAMs in extreme value analysis is well-established, particularly for modelling non-stationary processes. \citet{Chavez-Demoulin2005} applied additive models to threshold exceedance models, while \citet{Eastoe2009} extended these methods for covariate-dependent extreme value models. More recently, \citet{Youngman2019, Youngman2020} demonstrated the utility of GAMs for extreme value modelling and presented associated software, namely the \texttt{evgam} package.

We define a set, $Z$, of candidate covariates through which we aim to explain the non-stationarity of the process $M(m_t, y_t, \boldsymbol{s})$. Suppose that we have $K \in \mathbb{N}_+$ candidate covariates which we denote as $z_i(t, \boldsymbol{s}) \in Z$ where $i = 1, 2, \dots, K$, where each covariate is indexed by space, $\boldsymbol{s}$, and time, $t$. Note that, in the case of solely temporal covariates i.e., for the covariates $m_t$ and $A(y_t)$, $\boldsymbol{s}$ can be the whole spatial domain, $\mathcal{S}$, and, in the case of solely spatial covariates, i.e., $(Long, Lat)$ and $ c_{\sigma^2}(\boldsymbol{s})$, $t$ can be the entire temporal domain, $\mathcal{T}$. We assume that given a set of appropriate covariates, $Z$, we have that
\begin{align}
    \left\{M(m_t, y_t, \boldsymbol{s}) \mid Z\right\} \sim \text{GEV}\left\{\mu(t, \boldsymbol{s}), \sigma(t, \boldsymbol{s}), \xi(t, \boldsymbol{s}) \right\}.
\end{align}
Each parameter of the GEV can take an additive form, combining any or all covariates $z_i(t, \boldsymbol{s}) \in Z$.
The relationship between a covariate and each parameter of the GEV can be modelled linearly or with greater flexibility using a smoothing spline, which we denote as $\ell(\cdot)$. For example, a general formulation of the location parameter of the GEV can be expressed as
\begin{align}
    \mu(t, \boldsymbol{s}) = \gamma_0^{(\mu)} + \sum_{i=1}^{K} \gamma^{(\mu)}_i z_i(t, \boldsymbol{s}) + \sum_{i=1}^{K} \ell^{(\mu)}\{z_i(t, \boldsymbol{s})\},
\end{align}
where $\gamma^{(\mu)}_0$ is the intercept term and $\gamma^{(\mu)}_i$ represents the linear effect of covariate $z_i(t, \boldsymbol{s})$. The smoothing spline takes the form
\begin{align}
    \ell^{(\mu)}\{z_i(t, \boldsymbol{s})\} =  \sum_{j=1}^{P_i} \gamma^{(\mu)}_{ij} b^{(\mu)}_{ij}\left\{z_i(t, \boldsymbol{s})\right\}
\end{align}
where $P_i$ is the number of basis functions for covariate $z_i(t, \boldsymbol{s})$, with $b_{ij}(\cdot)$ being the spline basis functions and  $\gamma_{ij}$ the spline coefficients. We use penalised splines methods as implemented in \texttt{evgam}.

Furthermore, we allow random slopes for covariates to capture differential effects across space or time. For a given covariate $z_i(t, \boldsymbol{s})$, we introduce a random slope $\gamma^{(\mu)}_{i, t, \boldsymbol{s}} z_i(t, \boldsymbol{s})$, so that
\begin{align}\label{eq:general_gam_loc}
    \mu(t, \boldsymbol{s}) = \gamma_0^{(\mu)} +  \sum_{i=1}^{K} \gamma^{(\mu)}_i z_i(t, \boldsymbol{s}) + \sum_{i=1}^{K} \ell^{(\mu)}\{z_i(t, \boldsymbol{s})\} + \sum_{i=1}^{K} \gamma^{(\mu)}_{i, t, \boldsymbol{s}} z_i(t, \boldsymbol{s}).
\end{align}
Here, $\gamma^{(\mu)}_{i, t, \boldsymbol{s}}$ represents a time and location-specific deviation from the global effect of covariate $z_i(t, \boldsymbol{s})$. By including random effects, we improve the model's ability to capture latent spatial and temporal heterogeneity, leading to a more flexible and accurate characterisation of the extremal process. We also explore the inclusion of random intercepts for categorical covariates (such as month, $m_t$), however, we found this extension did not yield significant model improvement, so we do not detail the model formulation here. Models equivalent to that of Equation~\eqref{eq:general_gam_loc} can be written out for the log scale parameter, $\ln \sigma(t, \boldsymbol{s})$ and the shape parameter, $\xi(t, \boldsymbol{s})$, of the GEV.

To include higher-order interactions and cyclical effects, we consider tensor product splines, denoted $\ell_{\text{te}}(\cdot)$, and cyclic splines, denoted $\ell_{\text{cc}}(\cdot)$. Tensor product splines allow flexible interactions between covariates, while cyclic splines ensure continuity in periodic covariates (e.g., seasonal effects). These extensions are within the scope of the non-parametric model parameterisation we explored, providing additional flexibility in capturing complex dependencies and expanding the model space.

\subsection{Extremal dependence models}\label{sec:spatial_models}
The univariate GEV distribution arises as the limiting distribution of properly normalised block maxima from \textit{iid} sequences and is widely applied for statistical modelling of extreme events in one dimension, such as monthly maximum precipitation at a single location. However, environmental extreme events are not isolated. To model the joint behaviour of extremes across multiple sites, the theory must extend from a single distribution to a full stochastic process. This leads naturally to the study of max-stable processes, which are the infinite-dimensional generalisations of the GEV distribution \citep{deHaan1984}. Max-stable processes describe the limiting behaviour of component-wise maxima over collections of random fields, however, they are restrictive in practice, as they impose a strong form of extremal dependence (asymptotic dependence) which is typically not found in environmental data \citep{Huser2025}. 

When empirical evidence suggests that extremal dependence weakens at higher quantiles, asymptotically dependent models (such as the max-stable process or generalised Pareto process) are not suitable and risk overestimating extreme events. Hourly precipitation over the Piave basin exhibits asymptotic independence as shown in Section~\ref{sec:asymp_dep}. To accurately and realistically describe the behaviour of the extremal process, a model should capture the rate of decay of extremal dependence with increasing quantiles. Max-id models provide an extension to the well-studied max-stable model, maintaining the familiar methodological features of the max-stable theory while relaxing restrictions on the extremal dependence of the process. An attractive and practical feature of the relaxed assumptions on strong tail dependence is that we do not need to be certain that the block size is large enough to accurately assume that asymptotic results hold, which is a fundamental assumption in the max-stable modelling methodology. Instead, we can fit the model to smaller block sizes, making better use of data, resulting in less uncertain parameter estimates. Following \cite{Huser2021}, we take parametric forms of the max-id process which maintains the max-stable processes at the boundary of the parameter space. This allows the parameter estimates to inform the practitioner as to whether the data exhibit asymptotic dependence, and thus, if a max-stable process would be a more appropriate modelling framework.

A distribution function $G$ on $\mathbb{R}^p$ is said to be max-id if $G^v$ remains a valid distribution function for any $v> 0 \in \mathbb{R}$ \citep{brown1977, Resnick1987}. Intuitively, this means that $G$ describes the behaviour of the component-wise maximum of $n$ independent random variables, where $G = F^n$ for any $n = 1, 2, \dots$. To describe the monthly maxima of hourly precipitation over the spatial domain $\mathcal{S}$, we assume these maxima can be modelled using a max-id process. A max-id process extends the multivariate max-id distribution to the spatial setting. We restrict ourselves to this spatial context by fixing $p=2$. A process $Z(\boldsymbol{s}):\mathbb{R}^2\to \mathbb{R}$ is called a max-id process, if, given a collection of sites, say, $\boldsymbol{s}_1, \boldsymbol{s}_2, \dots, \boldsymbol{s}_k \in \mathcal{S}$, the finite-dimensional joint distribution of $Z(\boldsymbol{s})$ at those sites is $P\{Z(\boldsymbol{s}_1) \leqslant z_1, Z(\boldsymbol{s}_2)\leqslant z_2 \dots Z(\boldsymbol{s}_k)\leqslant z_k\} =G^v(z_1, z_2, \dots, z_k)$ where $G^v$ is a valid multivariate max-id distribution for any $v>0$.  

Assuming standard Fréchet margins, any multivariate extreme value distribution $G$ has the form $G(\boldsymbol{x}) = \exp\left\{ -V(\boldsymbol{x})\right\}$, where $V$ is called the exponent function. The exponent function $V$ has the form
\begin{align}\label{eq:exp_meansure_gen}
    V\left(\boldsymbol{y}\right)=p\int_{S_{p}}\left\{ \max _{i=1, \ldots, p}\left(\frac{\omega_i}{y_i}\right)\right\} h\left(\omega_1, \ldots, \omega_p\right) d\omega_1, \dots, d\omega_p,
\end{align}
for $\boldsymbol{y} = (y_1, \dots, y_p)\in\mathbb{R}^p_+$, where $h$ is called the spectral density and is a density function defined on the $(p-1)$-dimensional simplex. One possible framework for constructing max-id processes employs the spectral representation of max-stable processes introduced by \cite{deHaan1984}. A max-id process can be constructed as the pointwise maxima of random functions sampled from a Poisson point process with an appropriate function space \citep{Huser2021}. The spectral representation of the max-id process is,
\begin{align}\label{eq:spec_den}
    Z(\boldsymbol{s}) = \max_{i = 1,2,\dots}R_iW_i(\boldsymbol{s}),
\end{align}
where $\boldsymbol{s} \in \mathbb{R}^2$. By $W_i(\boldsymbol{s})$ we denote independent copies of the spatial process which describes the spatial characteristics of the extremal process $Z$ and $R_i$, a scalar, characterising the magnitude of the max-id process $Z$. We take $R_i$ for $i = 1,2,\dots$ to be realisations of a Poisson point process with mean measure described by $\kappa$. The dependence structure of the max-id process is completely characterised by the exponent measure of the Poisson point process. By defining the exponent measure appropriately, various extremal dependence structures, including asymptotic independence, can be effectively captured through careful specification of the mean measure in equation \eqref{eq:spec_den}. Notice that taking $\kappa = 1/r$ results in a process with Fréchet margins, i.e., the asymptotically dependent max-stable process. Choosing $\kappa$ to yield a lighter tail allows us to achieve an asymptotically independent model. Specifically, \cite{Huser2021} shows that taking
\begin{align}\label{eq:kappa}
    \kappa ([r, \infty]) = r^{-1}\exp\left\{ \beta^{-1} (1 - r^\beta) \right\},
\end{align} 
with $r>0$, corresponds to the Schlather max-stable model when $\beta = 0$. Taking $W(\boldsymbol{s}$) in equation \eqref{eq:spec_den} to be a Gaussian process, the max-id process, $Z(\boldsymbol{s})$ is Weibull-tailed and asymptotically independent when $\beta > 0$. Parameter estimates of $\beta \in [0, \infty)$ are particularly informative as they can describe the process's proximity to max-stability, i.e., an estimate $\hat \beta$, very close to zero suggests the process exhibits asymptotic dependence and so, alternative models are available that exploit the strong dependence between $R$ and $W$. Alternatively, larger values of $\beta$ indicate weaker dependence in the tails. To model the spatial profile in \eqref{eq:spec_den}, we take $W(\boldsymbol{s})$ to be a Gaussian process with the correlation function,
\begin{align}\label{eq:corr_func}
    \rho(d) = \exp\{-(d/\lambda)^\nu\},
\end{align}
where $d\in\mathbb{R}_+$ is the separation between a pair of sites, as defined in Section~\ref{sec:cov_dep}. While the parameter $\beta$, in \eqref{eq:kappa} describes the process's proximity to max stability, $\lambda \in [0, \infty]$ describes the spatial range of extremal dependence and $\nu\in(0,2]$, the smoothness parameter. 

We reparameterise the max-id process to account for seasonality in the extremal dependence structure. We account for within year variability using the covariate $T(m_t)$ as introduced in Section~\ref{sec:cov_dep}. We use this covariate to capture the fact that warmer temperatures in summer months increase atmospheric instability and moisture, leading to short-lived, localised convective storms. In contrast, colder winter temperatures produce large-scale frontal systems and widespread stratiform precipitation \citep{Eggert2015}. By using the covariate $T(m_t)$, we propose a model that can link extremal dependence structures to the physical temperature–precipitation relationships in the region. We allow for seasonality in the parameters of both $R$ and $W$ in the max-id process. For $R$, we incorporate seasonality by allowing the $\beta$ parameter of the mean measure of the Poisson point process to vary linearly with mean temperature, $T(m_t)$ (see Section~\ref{sec:cov_dep}). Larger values of $\beta$ would correspond to extreme events with a weaker form of extremal dependence, i.e., as events become more extreme, they become less spatially correlated. Allowing this parameter to change with monthly mean temperature allows the max-id model to reflect the varying extremal dependence levels of events throughout the year. Specifically, we let 
\begin{equation}
    \ln \beta(t) = \alpha_0^{(\beta)} + \alpha_1^{(\beta)} T(m_t).
\end{equation}
Additionally, for the spatial profile of the max-id process, we introduce seasonality by allowing the spatial range of dependence in the correlation function $\rho$, to vary linearly with $T(m_t)$. Larger values of $\rho$ correspond to more widespread extreme events. Allowing this parameter to change with monthly mean temperature allows the max-id model to reflect the varying spatial footprint of extreme events throughout the year. So, we take 
\begin{align}
    \ln \lambda(m_t) = \alpha_0^{(\lambda)} + \alpha_1^{(\lambda)}  T(m_t).
\end{align}
The temporally non-stationary max-id process is now,
\begin{equation}
    Z(t, \boldsymbol{s}) = \max_{i = 1,2,\dots} R_i(t) W_i(t, \boldsymbol{s}), ~~ \boldsymbol{s} \in \mathbb{R}^2.
\end{equation}

\subsubsection{Pairwise copula framework}
In order to fit the max-id model to the data, we use a copula framework. A copula is a multivariate distribution function with uniform marginal distributions on $[0, 1]$ \citep{Nelson2006}. Given a set of variables $\left(X_1, X_2, \dots, X_p\right)$, the copula function is the joint distribution of these variables after being transformed to uniform margins,
\begin{equation}
    C(u_1, u_2, \dots, u_p) = P\left\{ F_1(X_1) \leq u_1,  F_2(X_2) \leq u_2, \dots,  F_p(X_p) \leq u_p \right\},
\end{equation}
where $F_i$ is the marginal distribution function of $X_i$. If the margins are continuous, the copula function is unique \citep{Sklar1959} and invariant to marginal transformations. Copulas, therefore, make for a popular and widely used tool in multivariate extreme value analysis. The practitioner can first, separately model the marginal distributions of the process, transform the marginal distributions to be uniformly distributed using the probability integral transform \citep{Angus1994}, and subsequently model the extremal dependence. Any marginal transformation of a max-id vector remains max-id, therefore, the properties of max infinite divisibility lie within the dependence structure of the process. Thus, any copula which satisfies max-infinite divisibility can be used as a max-id model.

We write the exponent function of the max-id process in Equation~\eqref{eq:exp_meansure_gen}, on the uniform scale, as \begin{equation}
V^U(\boldsymbol{u})=V\left\{F_1^{-1}\left(u_1\right), \ldots, F_p^{-1}\left(u_p\right)\right\},
\end{equation}
where $F_i$ is the marginal distribution function of $X_i$ and $u_i$ denote data at site $s_i$ after being transformed to have uniform margins. The corresponding extreme value copula is then 
\begin{align}\label{eq:copula}
    C(\boldsymbol{u}) = \exp \left\{-V^U  (\boldsymbol{u}) \right\}.
\end{align}
In order to fit the max-id model in the copula framework, we transform the process $M(m_t, y_t, \boldsymbol{s})$ to have standard uniformly distributed marginals,  denoted as $M^U(m_t, y_t, \boldsymbol{s})$. To this end, we use the probability integral transform and the marginal models developed in Section~\ref{sec:marg_mod}. That is,
\begin{align}
    M^U(m_t, y_t, \boldsymbol{s}) = \hat F_{t, \boldsymbol{s}}\left\{M_U(m_t, y_t, \boldsymbol{s})\right\}
\end{align}
where $\hat F_{t, \boldsymbol{s}}(\cdot)$ is the estimated GEV distribution function of $M(m_t, y_t, \boldsymbol{s})$ at time $t$ and location $\boldsymbol{s}$. 

Recall that $\mathcal{S} = \{\boldsymbol{s}_1, \boldsymbol{s}_2, \dots, \boldsymbol{s}_p\}$, $p=49$, is the set of locations over which the max-id process is observed. Given $n$ observations of the max-id process, in our case the $n = 396 =12\times 33$ monthly spatial component-wise maxima from 1990-2023, the full likelihood function of the copula~\eqref{eq:copula} is 
\begin{equation}\label{eq:LL}
    \mathcal{L}\left(\boldsymbol{\theta} ; \boldsymbol{u}_1, \ldots, \boldsymbol{u}_n\right)=\prod_{i=1}^n\left[\exp \left\{-V^U\left(\boldsymbol{u}_i\right)\right\} 
    \sum_{B \subset \mathcal{S}} \prod_{\boldsymbol{s}_j \in B}\left\{-V_{B}^{U}\left(\boldsymbol{u}_{i, B} \right)\right\}\right]
\end{equation}
where the sum is over all non-empty subsets of $\mathcal{S}$ and $\boldsymbol{u}_{i, B}$ is the analogue of $\boldsymbol{u}_{i}$ for sites in $B$. $V_{B}^{U}$ denotes the $|B|$-th partial derivative of $V^{U}$ with respect to $\boldsymbol{u}_{\cdot, B}$. 

The cost of evaluating~\eqref{eq:LL} makes the maximisation problem computationally intractable. Recent advances in the literature have explored the use of machine learning methods to address similar issues for complex models for spatial extremes. The idea is to transfer the computational burden of fitting these models away from the practitioner, effectively amortising the cost of model fitting. Some proposed methods include variational autoencoders to emulate spatial extremes processes \citep{Zhang2024}, as well as Graph Neural Networks \citep{Sainsbury-Dale2025}. However, these estimation approaches are as yet only applicable to max-stable processes. Therefore, we instead rely on classical statistical inference approaches. In particular, we employ a composite likelihood approach \citep{Lindsay1988, cox2004}, namely an approximation based on combinations of pairwise likelihoods.
We calculate the bivariate log-likelihood between every pair of sites $\{\boldsymbol{s}_j, \boldsymbol{s}_k\} \subset \mathcal{S}$ to give the full composite log-likelihood, denoted $\mathcal{PLL}$, that is,

\begin{equation}
 \mathcal{PLL}\left(\boldsymbol{\theta} ; \boldsymbol{u}_1, \ldots, \boldsymbol{u}_n\right)=\sum\limits_{\substack{\{\boldsymbol{s}_j, \boldsymbol{s}_k\} \subset \mathcal{S} \\  j\neq k}}\ln \mathcal{L}\left\{\boldsymbol{\theta} ;\begin{pmatrix} u_{1 j} \\ u_{1 k} \end{pmatrix}, \ldots,\begin{pmatrix} u_{n j} \\ u_{n k} \end{pmatrix}\right\}.
\end{equation}
We are interested in the parameters $\boldsymbol{\theta}$ which maximise $\mathcal{PLL}$. We solve this problem via numerical optimisation to find the estimate
\begin{equation}
    \widehat{\boldsymbol{\theta}}=\underset{\theta}{\operatorname{argmax}}\mathcal{PLL} \left(\boldsymbol{\theta} ; \boldsymbol{u}_1, \ldots, \boldsymbol{u}_n\right).
\end{equation}
We found that taking all pairs led to the most stable parameter estimates however, additional computational efficiency can be achieved by taking a subset of pairs, as noted by \cite{Huser2021}
and by introducing a weighting of each pairwise likelihood based on the distance between sites. Statistical inference for max-id processes is computationally intensive, even more so when allowing for covariate-dependent parameters.

\section{Non-stationarities in extreme hourly precipitation over the Piave Basin}\label{sec:results}
In this section, we present the results of our case study on extremal precipitation in the Piave Basin. We begin with the cross-validation procedure used to assess marginal model performance in Section~\ref{sec:cross_val}, followed by a description of the marginal model development in Section~\ref{sec:gev_parameterisation}. Return levels derived from these models are reported in Section~\ref{sec:rl}. Finally, in Section~\ref{sec:res_extrem_dep} we present the fitted max-id dependence model, compare approaches for incorporating non-stationarity, and examine seasonality in the extremal dependence structure across the region.

\subsection{Cross-validation}\label{sec:cross_val}
\begin{figure}[ht]
    \centering
    \includegraphics[width=0.4\linewidth]{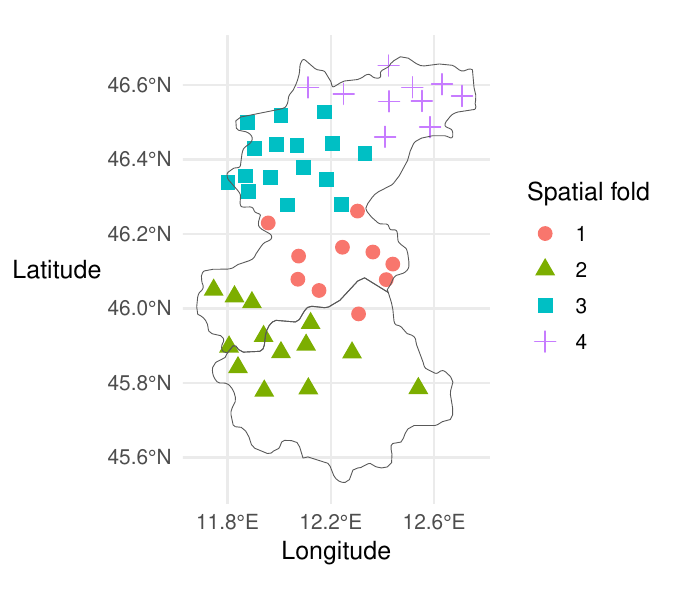}
    \caption{Spatial folds used for cross validation.}
\label{fig:spatial_folds}
\end{figure}

Given the richness of the proposed general marginal model structure in Equation~\ref{eq:general_gam_loc}, we require a model selection procedure, which we achieve through cross-validation. Our marginal model development follows a forward selection procedure. To compare models, we perform spatio-temporal cross-validation (CV) \citep[Chapter 7]{Hastie2008}, using normalised log-likelihood (nLL) and the continuously ranked probability score (CRPS) \cite{Zamo2018} as our CV scores. Unlike the standard choice of RMSE as a CV score, CRPS measures the difference between the predicted probability distribution function and observed values without needing to provide empirical quantiles, which are non-trivial to calculate in a spatio-temporally non-stationary setting. In each case, lower values of nLL and CRPS indicate a superior fit. We define $4$ spatial clusters over the region using the \texttt{r} package \texttt{spatialsample} \citep{Mahoney2023}, which are plotted in Figure~\ref{fig:spatial_folds}. We define $3$ temporal clusters, each consisting of every third month in the year, i.e., the first temporal cluster includes Jan, Apr, Jul and Oct. In this way, long-term temporal non-stationarity and seasonality are preserved in each of the three partitions. We define the 12 CV folds as the intersection of each unique pair of temporal and spatial clusters. In turn, we remove each fold from the dataset, fit the model to the remaining data and calculate the CV scores on the left-out fold. Finally, we take the average of the score over all 12 folds as the final score, which are presented in Table~\ref{table:tab_cv_gpd}. In the following section, we detail our forward selection approach to the parameterisation of our GEV model.

\subsection{Marginal model parameterisation}\label{sec:gev_parameterisation} 
A selection of the GEV model parameterisations we explored is presented in Table~\ref{table:tab_cv_gpd}. For each model, there is an intercept term $\beta_0^{(\mu)}$, $\beta_0^{(\sigma)}$ and $\beta_0^{(\xi)}$ for the location, scale and shape parameter respectively. For brevity, these terms are not displayed in the table. Model M0 acts as a base model, which is constant in space and time for each parameter of the GEV.

\newpage
\begingroup
\scriptsize 
\setlength{\LTleft}{-1.2cm}
\setlength{\LTright}{0pt}
\begin{longtable}{l|lllcc}
\caption{\normalsize Cross-validation metrics for each GEV model.}\label{table:tab_cv_gpd}\\ 
\textbf{M} & \textbf{$\mu$} & \textbf{$\ln \sigma$}&  \textbf{$\xi$} & \textbf{nLL} & \textbf{CRPS} \\
\hline\hline
  0 &  &  &  & -3.312 & 4.358\\  \hline
  1 & $ \gamma_1^{(\mu)} A(y_t)$ &  &   & -3.307 & 4.338\\  
  2 & $ \boldsymbol\gamma_{1,m_t}^{(\mu)} A(y_t)$ &  &   & -3.130 & 3.637\\  
  3 & $ \gamma_1^{(\mu)} A(y_t)+ \ell^{(\mu)}_{cc}(m_t) $ &  &   &  -3.092 & 3.529\\  \hline
  4 & $ \gamma_1^{(\mu)} D(\boldsymbol{s})$ &  &   & -3.303 & 4.276\\  
  5 & $ \gamma_1^{(\mu)} \text{Alt}(\boldsymbol{s})$ &  &   & -3.305 & 4.304\\ 
  6 & $ \ell^{(\mu)}_\text{te}(\text{Long}, \text{Lat})$ &  &   &  -3.300 & 4.251\\ 
  7 & $ \gamma_1^{(\mu)} c_\mu(m_t, \boldsymbol{s})$ &  &  & -3.286 & 4.178\\  \hline
  8 & $ \gamma_1^{(\mu)} A(y_t)+ \ell^{(\mu)}_{cc}(m_t) +  \gamma_2^{(\mu)}c_\mu(m_t, \boldsymbol{s})$ &  &  & -3.076 & 3.460\\ 
  9 & $ \gamma_1^{(\mu)} A(y_t)+ \ell^{(\mu)}_{cc}(m_t) +  \gamma_2^{(\mu)}c_\mu(m_t, \boldsymbol{s})$  & $\gamma_1^{(\sigma)} \ln c_{\sigma^2}(\boldsymbol{s}) $ &  & -3.070 & 3.432\\  
  10 & $ \gamma_1^{(\mu)} A(y_t)+ \ell^{(\mu)}_{cc}(m_t) + \gamma_2^{(\mu)}c_\mu(m_t, \boldsymbol{s})$  & $ \gamma_1^{(\sigma)} A(y_t) + \gamma_2^{(\sigma)} \ln c_{\sigma^2}(\boldsymbol{s}) $ &  & -3.068 & 3.430\\  
  11 & $ \gamma_1^{(\mu)} A(y_t)+ \ell^{(\mu)}_{cc}(m_t) +  \gamma_2^{(\mu)}c_\mu(m_t, \boldsymbol{s})$  & $ \gamma_{m(t),1}^{(\sigma)} A(y_t) + \gamma_2^{(\sigma)} \ln c_{\sigma^2}(\boldsymbol{s}) $ & & -3.018 & 3.312\\  
  12 & $ \gamma_1^{(\mu)} A(y_t)+ \ell^{(\mu)}_{cc}(m_t) +  \gamma_2^{(\mu)}c_\mu(m_t, \boldsymbol{s})$  & $ \ell_{cc}(m_t) + \gamma_2^{(\sigma)} \ln c_{\sigma^2}(\boldsymbol{s}) $ &  & -2.995 &3.303 \\  
  13& $ \gamma_1^{(\mu)} A(y_t)+ \ell^{(\mu)}_{cc}(m_t) +  \gamma_2^{(\mu)}c_\mu(m_t, \boldsymbol{s})$  & $ \gamma_1^{(\sigma)} A(y_t) + \ell_{cc}(m_t) + \gamma_2^{(\sigma)} \ln c_{\sigma^2}(\boldsymbol{s}) $ &  & -2.990 & 3.291\\  
  \hline
  14 & $ \gamma_1^{(\mu)} A(y_t)+ \ell^{(\mu)}_{cc}(m_t) +  \gamma_2^{(\mu)}c_\mu(m_t, \boldsymbol{s})$  & $ \gamma_1^{(\sigma)} A(y_t) + \ell^{(\sigma)}_{cc}(m_t) + \gamma_2^{(\sigma)} \ln c_{\sigma^2}(\boldsymbol{s}) $ & $\ell^{(\xi)}\{D(\boldsymbol{s})\}$ &  -2.984 & 3.239\\  
\rowcolor{green!20}  15 & $ \gamma_1^{(\mu)} A(y_t)+ \ell^{(\mu)}_{cc}(m_t) +  \gamma_2^{(\mu)}c_\mu(m_t, \boldsymbol{s})$  & $ \gamma_1^{(\sigma)} A(y_t) + \ell^{(\sigma)}_{cc}(m_t) + \gamma_2^{(\sigma)} \ln c_{\sigma^2}(\boldsymbol{s}) + \ell^{(\sigma)}\{D(\boldsymbol{s})\} $ & $\ell^{(\xi)}\{D(\boldsymbol{s})\}$ & { -2.922} & 3.100\\  
  16 & $ \gamma_1^{(\mu)} A(y_t)+ \ell^{(\mu)}_{cc}(m_t) +  \gamma_2^{(\mu)}c_\mu(m_t, \boldsymbol{s})$  & $ \gamma_1^{(\sigma)} A(y_t) + \ell^{(\sigma)}_{cc}(m_t) + \gamma_2^{(\sigma)} \ln c_{\sigma^2}(\boldsymbol{s}) + \ell^{(\sigma)}\{D(\boldsymbol{s})\} $ & $ \gamma_1^{(\xi)} A(y_t) + \ell^{(\xi)}\{D(\boldsymbol{s})\}$ &  -2.922 & 3.100\\  
  17 & $ \gamma_1^{(\mu)} A(y_t)+ \ell^{(\mu)}_{cc}(m_t) +  \gamma_2^{(\mu)}c_\mu(m_t, \boldsymbol{s})$  & $ \gamma_1^{(\sigma)} A(y_t) + \ell^{(\sigma)}_{cc}(m_t) + \gamma_2^{(\sigma)} \ln c_{\sigma^2}(\boldsymbol{s}) + \ell^{(\sigma)}\{D(\boldsymbol{s})\} $ & $ \ell^{(\xi)}_{cc}(m_t) + \ell^{(\xi)}\{D(\boldsymbol{s})\}$ & -2.920 & 3.097\\  
  18 & $ \gamma_1^{(\mu)} A(y_t)+ \ell^{(\mu)}_{cc}(m_t) +  \gamma_2^{(\mu)}c_\mu(m_t, \boldsymbol{s})$  & $ \gamma_1^{(\sigma)}A(y_t) + \ell^{(\sigma)}_{cc}(m_t) + \gamma_2^{(\sigma)} \ln c_{\sigma^2}(\boldsymbol{s}) + \ell^{(\sigma)}\{D(\boldsymbol{s})\} $ & $ \gamma_1^{(\xi)} A(y_t) + \ell^{(\xi)}_{cc}(m_t) + \ell^{(\xi)}\{D(\boldsymbol{s})\}$ & -2.920 & 3.097\\  \hline
\end{longtable}
\endgroup

Models M1\textendash M3 account for temporal non-stationarity in the location parameter, $\mu$, with the inclusion of linear terms incorporating global mean temperature, $A(t)$, as described in Section~\ref{sec:cov_marg}. In order to account for seasonal variation, we explore including a random slope in the coefficient of the temporal covariate $A(t)$ in the location parameter for each month. This random slope enables us to estimate the rate of change in precipitation extremes on a monthly basis, helping identify periods with the fastest and slowest rates of change in extremal precipitation. Both M1 and M2 models make an improvement over a temporally stationary location parameter (M0) in terms of both nLL and CRPS. Model M2 allows a random monthly slope for the temporal covariate, $A(t)$ for each month, $m_t$, achieving a substantial gain in performance over a non-seasonal model (M1). Alternatively, we include a spline over months. In model M3, we find seasonality more effectively and parsimoniously incorporated through a separate cyclical spline over the months, $\ell_{cc}(m_t)$, with 12 knots, one placed at each month.

Models M4\textendash M7 investigate the efficacy of several spatial covariates to be included in the location parameter. These include a linear effect with coastal distance and altitude in models M4 and M5, respectively and a 2-dimensional spline over latitude and longitude in model M5. Model M7 uses the climate model-derived mean monthly maxima, $c_\mu$, as described in Section~\ref{sec:cov_marg}, which appears to be the most informative spatial covariate for the location parameter. 

Models M8\textendash M13 explore parameterisations of the log of the scale parameter, $\ln \sigma$. In model M8 (and all subsequent models), for the location parameter, we take the most effective representation of temporal and spatial non-stationarity from previous models. Model M9 incorporates the climate model-derived spatial covariate, namely, the empirically estimated variance of monthly maxima, $c_{\sigma^2}$, as described in Section~\ref{sec:cov_marg}. Models M10\textendash M13 build upon this by including temporal and seasonal terms. The model which yields the best performance in both nLL and CRPS is model M13, mirroring the parameterisation of the location parameter, i.e., a linear temporal term as well as a smoothing cyclical spline over the months. 

Models M14\textendash M18 allow for spatial and temporal non-stationarity in the shape parameter, $\xi$. A smoothing spline applied to coastal distance proved to be an effective spatial covariate for the shape parameter, substantially improving upon the CV scores of the equivalent model with constant shape (in model M13).

\begin{figure}[ht]
    \centering
    \includegraphics[width=\linewidth]{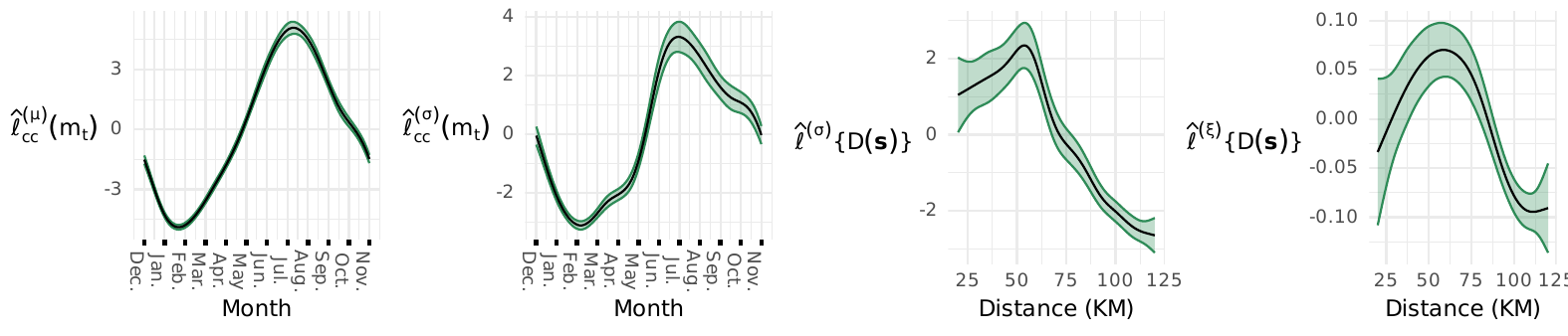}
    \caption{Estimated smooth terms for marginal model M15. From left to right: location and scale (months); scale and shape (coastal distance)}
    \label{fig:splines}
\end{figure}

Model M15 finds further substantial improvement by incorporating a smoothing spline on coastal distance into the scale parameter. This suggests that the empirically estimated descriptor of variance of monthly maxima does not fully capture the variability of the scale parameter fitted to the observed maxima. Model M16 accounts for temporal non-stationarity in the shape parameter. This results in a slight decrease in performance as compared to its temporally stationary competitor (M15) in terms of log-likelihood and no improvement in terms of CRPS. Models M17 - M18 attempt to account for seasonality in the shape parameter, yielding a very slight improvement in CV metrics. 
However, we note that the coefficient $\beta^{(\xi)}_1$ in both models M16 and M18 is not statistically significant, with the 95\% confidence interval containing $0$, and so, we do not prefer these models. It appears that seasonality and temporal non-stationarity of the process are amply captured by the other parameters of the GEV. There is a very slight improvement achieved by model M17 over M15; however, given the lack of clear or convincing superior performance achieved, we decide it is not worth the inclusion of the additional spline model, and choose the simpler temporally constant but spatially varying shape parameter, i.e, choosing model M15. This finding is in standing with the current literature \citep{Katz2002}, where it is well documented that estimation of the shape parameter in most environmental applications is challenging, and thus, in general, a simpler parameterisation of the parameter is preferred. The estimated smooth terms from model M15 are plotted in Figure~\ref{fig:splines}. In this figure, a clear seasonal pattern is captured by the smoothing splines $\hat\ell^{(\mu)}_{cc}(m_t)$ and $\hat\ell^{(\sigma)}_{cc}(m_t)$, for the location and scale parameter respectively. The splines suggest more intense and more variable extreme hourly precipitation in summer months, reflecting observed patterns identified in Section~\ref{sec:explor}. The estimated smoothing splines $\hat\ell^{(\mu)}_{cc}\left\{D(\boldsymbol{s})\right\}$ and $\hat\ell^{(\sigma)}_{cc}\left\{D(\boldsymbol{s})\right\}$ capture the effect of coastal distance for the scale and shape parameter respectively. There is a marked change in the behaviour of these parameters with respect to coastal distance at about 50 km+ inland. At this distance, it is clear that the Prealps begin to have a substantial impact on the behaviour of extreme precipitation.

\begin{figure}[ht]
    \centering
    \includegraphics[width=0.8\linewidth]{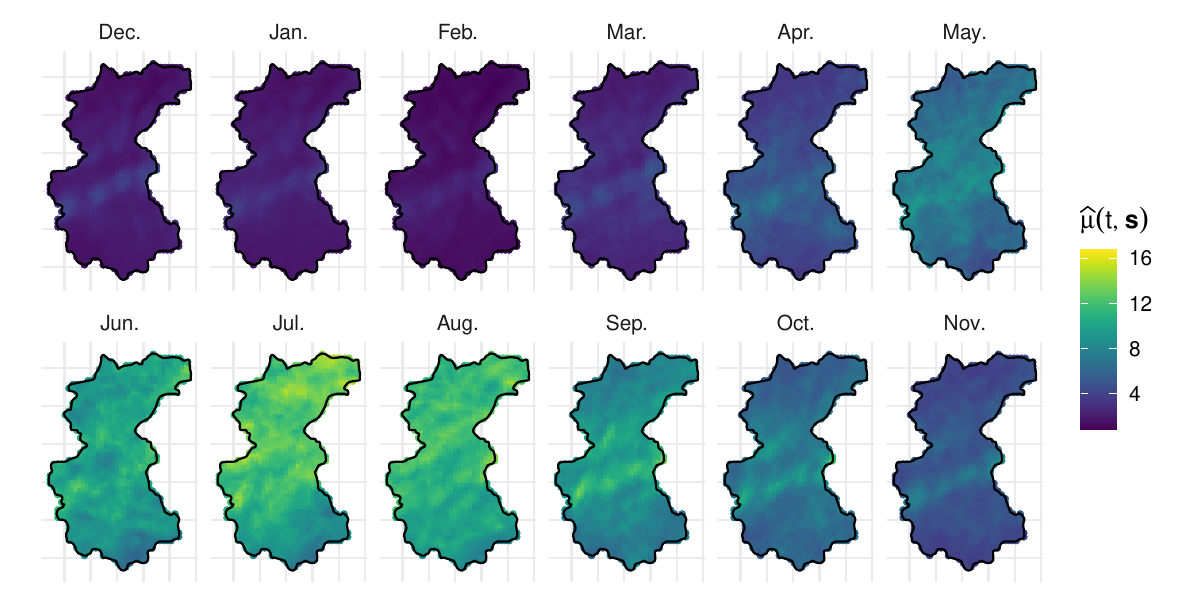}
    \caption{Estimate of the location parameter, $\hat\mu(t, \boldsymbol{s})$ from model M15 for each month, in the climatic context of $y_t = 2023$.}
    \label{fig:loc_est}
\end{figure}

\begin{figure}[ht]
    \centering
    \includegraphics[width=0.8\linewidth]{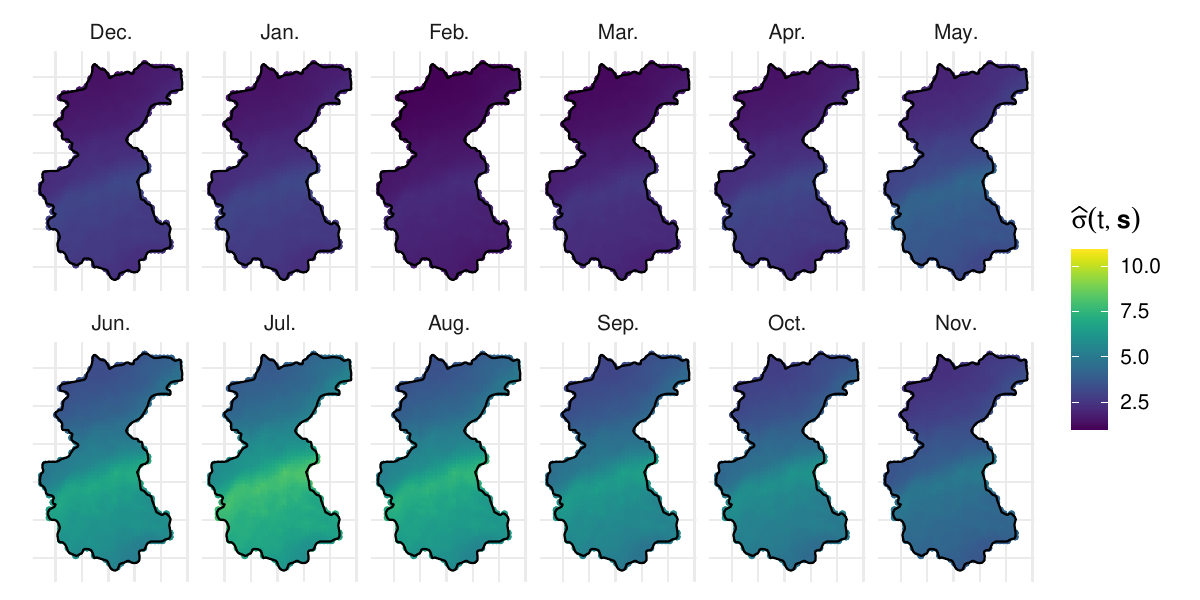}
    \caption{Estimate of the scale parameter, $\hat\sigma(t, \boldsymbol{s})$ from model M15 for each month, in the climatic context of $y_t = 2023$.}
    \label{fig:scale_est}
\end{figure}

\begin{figure}[ht]
    \centering
\includegraphics[width=0.23\linewidth]{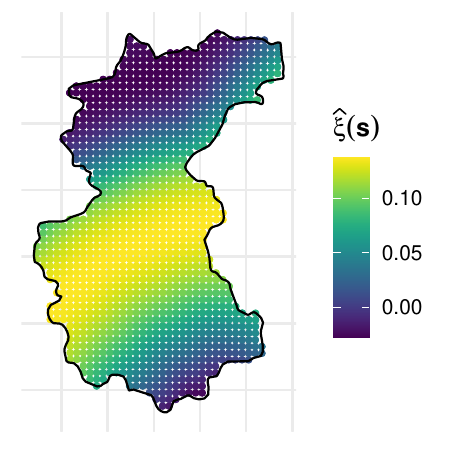}
    \caption{Estimate of the shape parameter, $\hat\xi(\boldsymbol{s})$ from model M15.}
    \label{fig:shape_est}
\end{figure}

The parameter estimates of the shape, location and scale parameters from model M15 are shown in Figures~\ref{fig:loc_est}, \ref{fig:scale_est} and \ref{fig:shape_est} respectively. The estimated location parameter, $\hat{\mu}(t, \boldsymbol{s})$, as shown in Figure~\ref{fig:loc_est}, inherits its spatial structure directly from the climate model output, allowing the model to capture fine-scale geophysical features and local climatic influences that could not be inferred from observational data alone. The estimated scale parameter, $\hat{\sigma}(t, \boldsymbol{s})$, as shown in Figure~\ref{fig:scale_est}, indicates generally higher variability in extreme precipitation in the mountainous regions. In contrast, lower scale values are observed in the lower-elevation plains of the region. Seasonally, the scale tends to increase during the summer months, consistent with the more convective and spatially heterogeneous precipitation processes that dominate during this period. Finally, the estimated shape parameter, $\hat\xi(t, \boldsymbol{s})$, as shown in Figure~\ref{fig:shape_est}, which governs the tail heaviness of the distribution, is highest at Prealps, highlighting the orographic lifting effect in this region. 

To further assess the performance of our chosen GEV model, quantile-quantile (QQ) plots for a selection of sites, both in and out-of-sample, are shown in Section~1 of the Supplementary Material, indicating a very good fit of the model.

\subsection{Return levels}\label{sec:rl}
\begin{figure}[ht]
    \centering
    \includegraphics[width=0.8\linewidth]{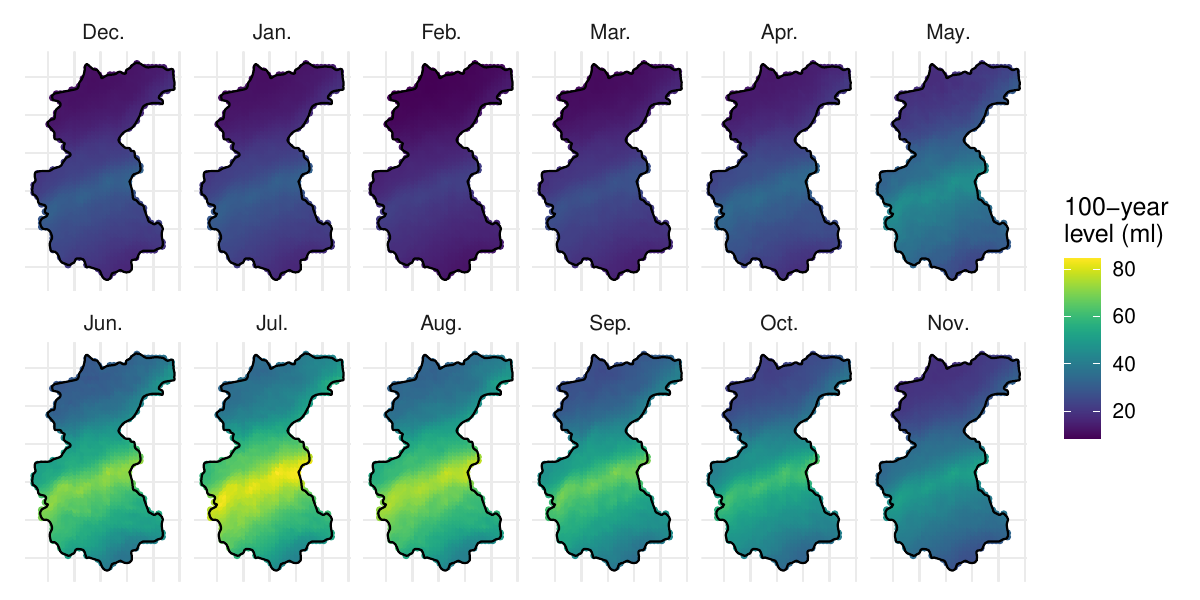}
    \caption{Estimated effective monthly 100-year return level of hourly precipitation for covariate $A(y_t) : y_t = 2023$.}
    \label{fig:rl_hourly_max} 
\end{figure}

\begin{figure}[ht]
    \centering
    \includegraphics[width=0.9\linewidth]{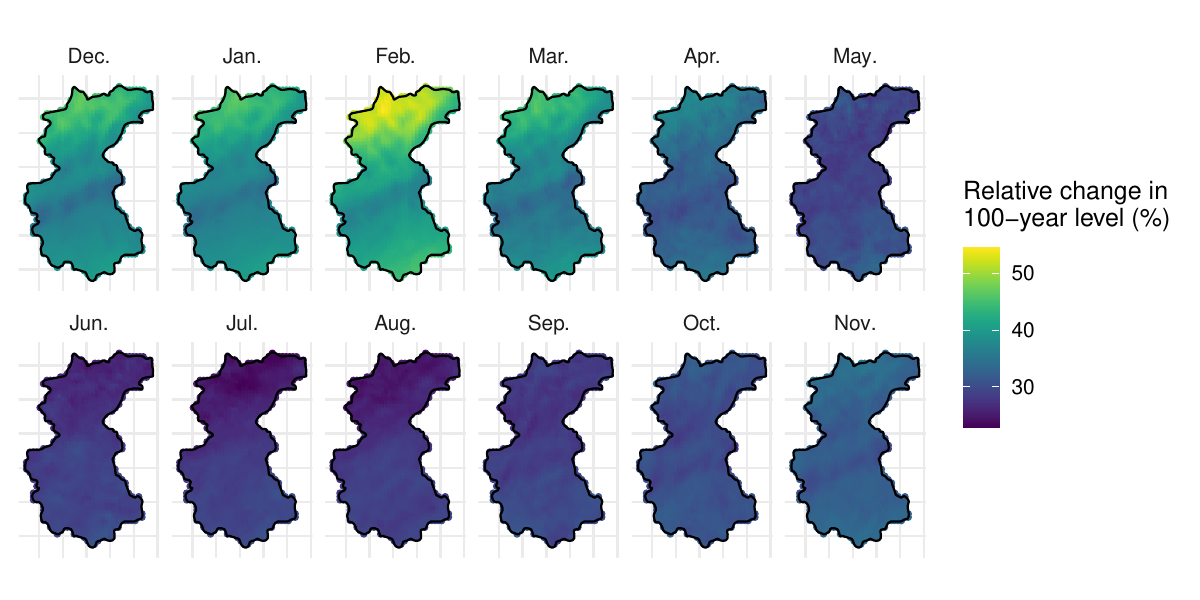}
    \caption{Relative change (\%) in 100-year return level of hourly precipitation data calculated for each month separately for $y_t = 2023$ with respect to $y_t = 1990$.}
    \label{fig:rl_rel_change_hourly_max}
\end{figure}

We estimate the effective 100-year return level associated with hourly precipitation for each month \citep{Katz2002}, as shown in Figure~\ref{fig:rl_hourly_max}. In Figure~\ref{fig:rl_rel_change_hourly_max}, we also plot the relative change in return level as a percentage increase from 1990 to 2023 for each month separately. Spatial variability of the estimated return levels reflects the heterogeneity of the climate dynamics acting over the study region, which have been informed by the underlying climate model output. A distinct transition from the flat plains in the south of the region to the start of the alpine regions is clearly captured and marked by a band of high return levels. That characteristic is especially pronounced during the late summer and early autumn periods, corresponding to the seasonality of atmospheric conditions that are favourable for the enhancement of convective activity and orographic lifting processes. The topographic lifting of precipitation extremes, associated with the lifting of moist air masses over elevated terrain, leads to higher return levels along the mountainous areas. The lowest return levels are confined to the northwestern part of the study domain. The relatively lower extreme precipitation over this region is likely due to the influence of shielding/shadowing by local topographic features \citep{Isotta2014}. By incorporating covariates derived from climate model such as $c_\mu(m_t, \boldsymbol{s})$ and $c_\sigma^2(\boldsymbol{s})$ our approach significantly enriches the spatial information in our return level maps, providing a much more detailed picture than would be possible using solely the gauged data. The model encapsulates the interplay of topographic features and extreme precipitation processes, which can be seen in the spatially varying return levels. These results illustrate the dominating role that local topography and seasonal atmospheric dynamics play in shaping extreme precipitation behaviour in the study region. 

\newpage
\subsection{Seasonal extremal dependence}\label{sec:res_extrem_dep}
To facilitate a faster computation time, we fix the shape in equation~\eqref{eq:corr_func} to $1$. For model comparison, we use both the log-likelihood as well as the composite likelihood information criterion \citep[CLIC]{Davison2012}, $\text{CLIC} = -2\mathcal{PLL}\left(\hat{\boldsymbol{\theta}}\right) + 2\text{tr}\left\{ J^{-1}\left(\hat{\boldsymbol{\theta}}\right) K\left(\hat{\boldsymbol{\theta}}\right) \right\}$, where $\mathcal{PLL}\left(\hat{\boldsymbol{\theta}}\right)$ is the composite pairwise log-likelihood of the estimated parameters, $J$ is the estimated hessian matrix of  $-\mathcal{PLL}\left(\hat{\boldsymbol{\theta}}\right)$, and $K$ the estimated variance of the score contributions. Both $J$ and $K$ are found through numerical derivations. Models with lower CLIC values indicate a better fit. To estimate the uncertainty of parameter estimates of the max-id model, we use the Hessian matrix. A full uncertainty assessment involving a bootstrap procedure would, at present, require access to very substantial computational resources.

We fit the max-id model described in Section~\ref{sec:spatial_models} to investigate and account for significant seasonality in the extremal dependence parameter $\beta$ and spatial correlation parameter $\lambda$. We found that extremal dependence of precipitation in the Piave basin is non-stationary. This is evidenced by substantial improvements in both LL and CLIC scores in Table~\ref{tab:maxidres}. We also present CLIC scores for the out-of-sample sites in the Supplementary Material, Table 1. Incorporating seasonality into the max-id model by means of the monthly temperature information significantly improves the model fit over the stationary model. These results suggest that, for the study of extremal precipitation, it is of great importance to consider seasonality. The extreme precipitation events in the Piave basin are influenced both by the seasonal changes in the atmospheric processes and by the topographical characteristics of the region, and those which are not sufficiently captured at the marginal level. 

\begin{longtable}{lccccr}
\caption{Parameter estimates and performance metrics of max-id process models.}
\label{tab:maxidres}\\

\toprule
\textbf{Model} & $\alpha_0^{(\beta)}$ & $\alpha_1^{(\beta)}$ & $\alpha_0^{(\lambda)}$ & $\alpha_1^{(\lambda)}$ & \textbf{CLIC} \\
\midrule
\endfirsthead

\toprule
\textbf{Model} & $\alpha_0^{(\beta)}$ & $\alpha_1^{(\beta)}$ & $\alpha_0^{(\lambda)}$ & $\alpha_1^{(\lambda)}$ & \textbf{CLIC} \\
\midrule
\endhead

\midrule
\multicolumn{6}{r}{\textit{Continued on next page}} \\
\midrule
\endfoot

\bottomrule
\endlastfoot

Stationary 
& 0.501 (0.499, 0.503)
& --
& 0.187 (0.183, 0.190)
& --
& -76,884 \\

Seasonal
& 0.673 (0.669, 0.678)
& -1.209 (-1.220, -1.197)
& -0.526 (-0.538, -0.513)
& 1.108 (1.102, 1.114)
& -108,833 \\

\end{longtable}

\begin{figure}[ht]
    \centering   \includegraphics[width=\linewidth]{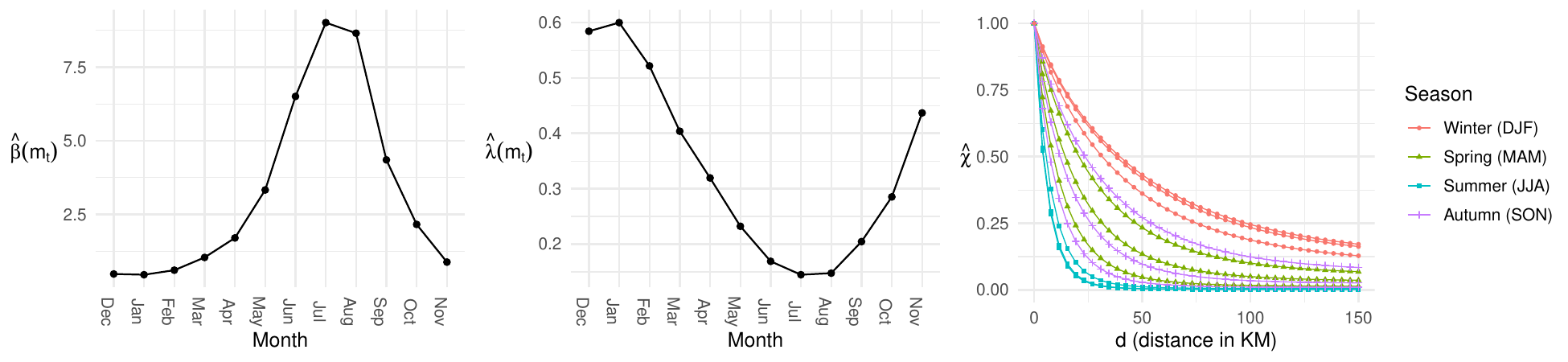}
    \caption{Parameter estimates of the max-id process. (left) estimate of $\beta$ for each month, (middle) estimate of parameter $\lambda$ for each month and (right) resulting estimates of pairwise extremal dependence coefficient $\hat \chi$ for all months, coloured by season to show seasonality captured by the max id model.}
    \label{fig:maxid_param_estimtes}
\end{figure}

We explore the inclusion of seasonality in the extremal dependence parameter $\beta$, which describes the ``proximity to max-stability" of extremal dependence between extreme events at different locations. As seen in Figure~\ref{fig:maxid_param_estimtes}, a clear seasonal pattern occurs in the parameter estimates. We estimate lower values of $\beta$ in winter months and higher values in summer months. The lower values of $\beta$ indicate stronger extremal dependence between sites at the same distance. This suggests a higher likelihood that extreme events co-occur at multiple sites. Conversely, higher values of $\beta$ indicate a weaker extremal dependence, i.e., co-occurring extremes are less likely and spatial coherence diminishes. Additionally, we found evidence of seasonality in the spatial range parameter, $\lambda$, of extreme hourly precipitation events, with the range being lower in summer months and higher in winter months, as shown in Figure~\ref{fig:maxid_param_estimtes}. The lower values of $\lambda$ during summer months suggest that extreme precipitation events are more localised during this period. These findings are in line with the dominance of convective processes in summer, where thunderstorms and localised convective cells often result in intense, short-duration rainfall events. These events are more isolated and spatially confined due to the localised nature of the convective storms, which usually have a limited spatial footprint \citep{Casallas2023}. On the contrary, larger values of $\lambda$ and lower values of $\beta$ in winter confirm that the extreme events at different precipitation events are spatially more coherent during the cold period. This is likely related to large-scale atmospheric circulation patterns typical of the winter season, also considering the orographic enhancement processes involving moist air masses intersecting with the mountain relief of the Piave basin \citep{Iannuccilli2021}. The larger-scale processes associated with frontal systems and more persistent low-pressure systems tend to produce the more spatially extensive extreme precipitation events, which is reflected in our parameter estimates for winter months \citep{Couto2024}. 

\begin{figure}[ht]
    \centering
\includegraphics[width=\linewidth]{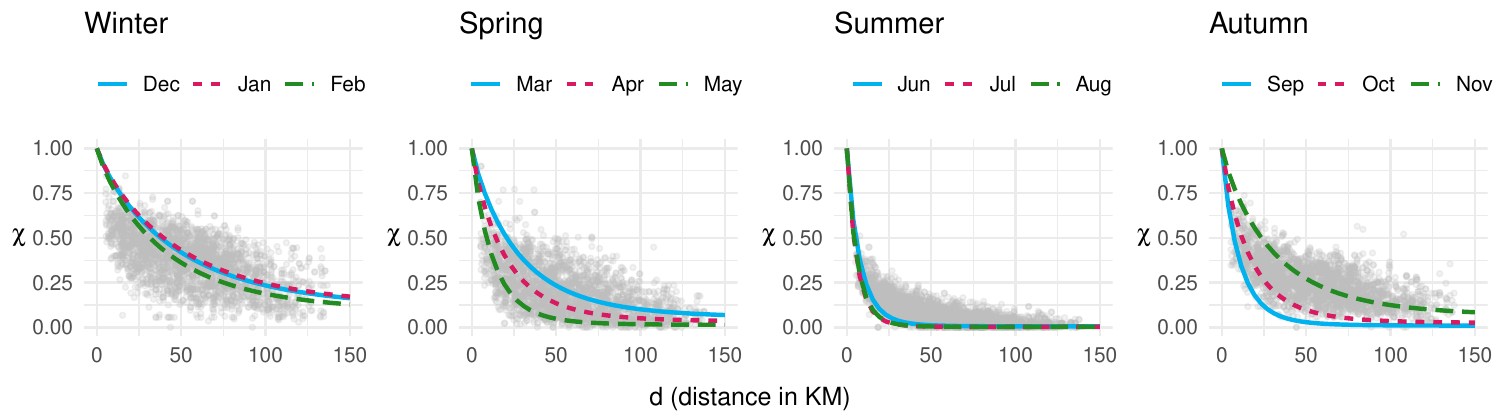}
    \caption{Estimates of $\chi$ against topographically adjusted inter-site distance $d$ derived from max-id model. A pairwise empirical estimator with $q = 0.997$ is shown as point clouds. Lines display the limiting value for $\chi$ estimated using the max-id process, with colour and line type corresponding to different months within each faceted season.}
    \label{fig:chi_res_maxid}
\end{figure}

In the right-hand plot of  Figure~\ref{fig:maxid_param_estimtes}, we derive estimates of $\chi$ from our fitted max-id model from each month separately (coloured by season), which combines the information contained in the parameters $\beta$ and $\lambda$. We can see a fast decay of extremal dependence with increased distance for summer months and the more persistent extremal dependence (even at substantial distances) for winter months. To asses how the max-id model fits the data and how it captures seasonality, in Figure~\ref{fig:chi_res_maxid}, we plot a point cloud of pairwise empirically estimated values of $\chi_{q = 0.997}$ against topographically adjusted inter-site distance $d$. The lines correspond to estimates derived from the max-id model. The figure shows a good fit of the model to the data, showing a good model fit. We investigated versions of the model in which either of the parameters $\lambda$ or $\beta$ is constant. However, we found this led to numerical instability.

\section{Discussion}\label{sec:discussion} 
This study provides a detailed analysis of the spatio-temporal characteristics of extreme hourly precipitation over the Piave Basin in northern Italy. We propose a suite of models which exploit recent methodological advances in extreme value analysis to account for the pronounced spatial heterogeneity and strong seasonal variability characteristic of extreme precipitation in this alpine basin. 
Accurate long-term return level estimates are critical for accurate risk assessment and the design of resilient infrastructures. This is of paramount importance to support the development of climate change adaptation strategies. We derive spatially rich return level estimates with a high level of physical detail, made possible by the assimilation of climate model outputs and observed weather data.

The specification of a parsimonious statistical model to capture the interplay of numerous sources of variability is challenging, and so, traditional statistical methods are poorly suited to this end. Our flexible, semi-parametric modelling framework addresses this challenge by allowing the inclusion of physically interpretable covariates. The identification of suitable predictors is a key contribution of our proposal. We propose local-scale predictors (e.g., coastal distance) and large-scale climate-driven variables, which can also capture the effect of climate change. This methodology allows us to reflect both local and large-scale variabilities. While projections are not within the scope of this research, incorporating covariates such as temperature anomalies provides a basis for projecting extreme precipitation events under different climate change scenarios. 

The flexibility of our model goes beyond marginal heterogeneities to capture the complex seasonal behaviour of spatial dependence for extremes, by exploiting the rich feature of the max-id processes. We find evidence to suggest that at higher intensities, extremal dependence of precipitation decreases. While traditional max-stable models assume asymptotic dependence, our results support the use of more flexible max-id processes, which allow for both asymptotic dependence and independence regimes. This flexibility is particularly crucial in mountainous regions where precipitation dynamics can vary drastically over short distances. We show that topographical distance, which incorporates elevation changes, gives a more physically realistic description of their separation as compared to a Euclidean or Haversine distance, which can be unsuitable in highly diverse topographical settings, where elevation-induced orographic effects significantly influence precipitation patterns. 

The introduction of seasonal variation into the parameters of the max-id model captures non-stationarity in both the extremal dependence levels and spatial range of extreme events. Our approach provides a clear physical interpretation and reveals a shift in extremal dependence behaviour of hourly precipitation extremes across different times of the year. Estimates of the max-id parameters reflect the stark differences in the climatic drivers of precipitation extremes in the region at different periods of the year, which arise from fundamentally different atmospheric processes.  The stronger extremal dependence observed in winter is consistent with widespread synoptic systems that impact large areas simultaneously. In contrast, weaker extremal dependence in summer corresponds to convective systems, which are more spatially isolated and produce localised extremes. Our model also captures the transition between these dominating precipitation regimes during spring and autumn, where the intermediate parameter values reflect the mixed atmospheric processes during those periods. We conclude that the heterogeneous behaviour of precipitation extremes cannot be fully accounted for by modelling non-stationarity only at the marginal level of the process. Non-stationarities of the extremal dependence must also be modelled. In our model, we included only one covariate to account for non-stationarity in the extremal dependence structure. Extending the model to include other meteorological variables such as humidity or wind patterns could improve representation of the underlying processes. We highlight this as a potentially insightful avenue for future work, although this would likely require more data and would be challenging from a computational point of view.


Our model was primarily developed to address hourly precipitation extremes, which are considered especially relevant for flood early-warning systems and urban hydrology applications, where short-duration intensities are most critical. However, the methodology can naturally be applied to model precipitation accumulated over other durations. As highlighted in recent studies \citep{Formetta2022}, different accumulation periods may require bespoke modelling strategies due to differences in storm dynamics and runoff response. Within the family of models we propose, such differences could be captured simply through the introduction and selection of adequate covariates for both the marginal and dependence structures. Furthermore, the applicability of the model is not limited to precipitation data, but extends to other types of spatial extremes.

\section*{Data and software availability}
Data are freely available from \url{https://www.arpa.veneto.it/dati-ambientali/dati-storici/meteo-idro-nivo/meteo-idro-dati-orari}. We thank Prof.~Enrico Bertuzzo for facilitating access to the data and for fruitful discussions about the hydrology of the basin. All code used in this analysis is available on GitHub\footnote{\url{https://github.com/dairer/Extreme-Precipitation-Piave-Basin}}.

\section*{Acknowledgments}
This study was carried out within the RISE project and received funding from the European Union Next-GenerationEU - National Recovery and Resilience Plan (NRRP) – MISSION 4 COMPONENT 2, INVESTIMENT 1.1 Fondo per il Programma Nazionale di Ricerca e Progetti di Rilevante Interesse Nazionale (PRIN) – CUP N.H53D23002010006. This publication reflects only the authors’ views and opinions; neither the European Union nor the European Commission can be considered responsible for them. We thank Professor Jonathan A. Tawn for helpful discussion.

\bibliography{refs}  






\end{document}